\documentclass[12pt,preprint]{aastex}

\begin{document}

\title{A possible giant planet orbiting the cataclysmic variable LX Ser}

\author{Li K.\altaffilmark{1,2}, Hu, S.-M.\altaffilmark{1}, Zhou, J.-L.\altaffilmark{3}, Wu, D.-H.\altaffilmark{3}, Guo, D.-F.\altaffilmark{1}, Jiang, Y.-G.\altaffilmark{1}, Gao, D.-Y.\altaffilmark{1}, Chen, X.\altaffilmark{1}, Wang, X.-Y.\altaffilmark{1}}

\altaffiltext{1}{Shandong Provincial Key Laboratory of Optical Astronomy and Solar-Terrestrial Environment, Institute of Space Sciences, Shandong University, Weihai, 264209, China
 (e-mail: kaili@sdu.edu.cn, likai@ynao.ac.cn (Li, K.), husm@sdu.edu.cn (Hu, S.-M.))}
\altaffiltext{2}{Key Laboratory for the Structure and Evolution of Celestial Objects, Chinese Academy of Sciences, Kunming 650011, China}
\altaffiltext{3}{School of Astronomy and Space Science and Key Laboratory of Modern Astronomy and Astrophysics in Ministry of Education, Nanjing University, Nanjing 210093, China}

\begin{abstract}
LX Ser is a deeply eclipsing cataclysmic variable with an orbital period of $0.^d 1584325$. Sixty two new eclipse times were determined by our observations and the AAVSO International Data base. Combining all available eclipse times, we analyzed the $O-C$ behavior of LX Ser. We found that the $O-C$ diagram of LX Ser shows a sinusoidal oscillation with a period of 22.8 yr and an amplitude of 0.00035 days. Two mechanisms (i.e., the Applegate mechanism and the light travel time effect) are applied to explain the cyclic modulation. We found that the Applegate mechanism is difficult to explain the cyclic oscillation in the orbital period. Therefore, the cyclic period change is most likely to be caused by the light travel time effect due to the presence of a third body. The mass of the tertiary component was determined to be $M_3\sim7.5 \,M_{Jup}$. We supposed that the tertiary companion is plausible a giant planet.
The stability of the giant planet was checked, and we found that the multiple system is stable.

\end{abstract}


\section{Introduction}
Cataclysmic variables (CVs) are interacting binary stars composed of a degenerate white dwarf primary and a Roche lobe filling
M dwarf secondary. LX Ser was first identified to be an eclipsing CV by Stepanian (1979) when searching galaxies with ultraviolet continuum and was classified to be SW Sex subclass of CVs. Stepanian (1979) found that the brightness of LX Ser was about $m_V=14$ mag. Recently, the orbital inclination and mass ratio of LX Ser were determined to be $i=79^\circ.0$ and $q=0.50$ by Marin et al. (2007). By analyzing the eclipse times of LX Ser, Horne (1980) determined an ephemeris $HJD=2444293.02377(\pm0.00020)+0^d.1584328(\pm0.0000010)E$ and showed that an upper limit on \.{P} is $10^{-5}$. Eason et al. (1984) derived a new ephemeris using a second-order least-squares fitting of all the eclipse times, and the second-order term was insignificant and was ignored.  After that, many new eclipse times were obtained (e.g., Agerer \& Hubscher 2003; Diethelm 2003; Zejda 2004; Hubscher et al. 2005; Krajci 2005, 2006; Zejda et al. 2006), allowing us to reanalyze the eclipse times.

Eclipsing times of CVs can be determined with high precision, and small amplitude orbital period changes could be discovered. Therefore, CVs are very good targets to search for planetary-mass companions. The presence of the planetary companions orbiting the central stars will cause small wobbles of the barycenter of the triple system. The light from the stars will travel closer or further due to the variation of the distance between the host system and the earth. Then, the arrival eclipse times vary cyclically and the observed-calculated ($O-C$) diagram shows a cyclic change. By analyzing the $O-C$ variations, very small mass companions can be detected. This method has been successfully used to detect substellar companions surrounding CVs, e.g., Z Cha (Dai et al. 2009), DP Leo (Qian et al. 2010a), QS Vir (Qian et al. 2010b), UZ For (Potter et al. 2011), V2051 Oph (Qian et al. 2015). In this paper, we show the investigation of the eclipse times of LX Ser and the possible giant planet orbiting this system.

\section{New Observations}
New photometric observations of LX Ser were carried out on April 24, May 30, and June 26, 2015, and February 20, 2016. The CCD images were taken by using the 1.0 m Cassegrain telescope at Weihai Observatory of Shandong University (Hu et al. 2014) with the Andor DZ936 camera. The size of each pixel is 0.35$\arcsec$, resulting that the effective field of view is about 11.8$'$ $\times$ 11.8$'$. The observation information is listed in Table 1. All the CCD images were analyzed using the IMRED and PHOT packages in IRAF\nolinebreak\footnotemark[1] procedure\footnotetext[1]{IRAF is distributed by the National Optical Astronomy Observatories, which is operated by the Association of Universities for Research in Astronomy Inc., under contract to the National Science Foundation.}. The light curve in $R_c$ band observed on April 24, 2015 is displayed in Figure 1 for example. The brightness flickering outside the eclipse can be seen, which is caused by variations of mass transfer rate from the secondary red dwarf to the primary white dwarf. Four new times of light minimum were determined and are shown in Table 2. Our new eclipse times are 3 or 4 significant figures, indicating very high precision. We also collected the data from AAVSO (American Association of Variable Stars Observers) International Data base\footnote{http://www.aavso.org/}. Based on the data, 58 eclipse times were reanalyzed and are also listed in Table 2, the corresponding light curves are displayed in Figure 2. All the eclipse times were determined by this work using the parabolic fitting method and was converted to Barycentric Julian Dates (BJD) using the software of Eastman et al. (2010). As seen in Figures 1 and 2, all the eclipse times can be fitted by parabolic curves very well.

\begin{table}
\begin{center}
\caption{Observation information of LX Ser}
\begin{tabular}{lccc}\hline\hline
Date	         &Duration time (hr)&	Filter&	Exposure time (s)\\  \hline
24 Apr., 2015	 & 3.89	            & R$_c$	  & 100              \\
30 May, 2015	 & 0.78	            & R$_c$	  & 100              \\
26 Jun., 2015	 & 0.65	            & V	    & 100              \\
20 Feb., 2016	 & 0.94	            & N$^1$	    & 30               \\\hline
\end{tabular}
\end{center}
\scriptsize $^1$ N means no filter.
\end{table}

\begin{figure}
\begin{center}
\begin{tabular}{c@{\hspace{0pc}}c}
\includegraphics[width=8.0cm,height=7.0cm]{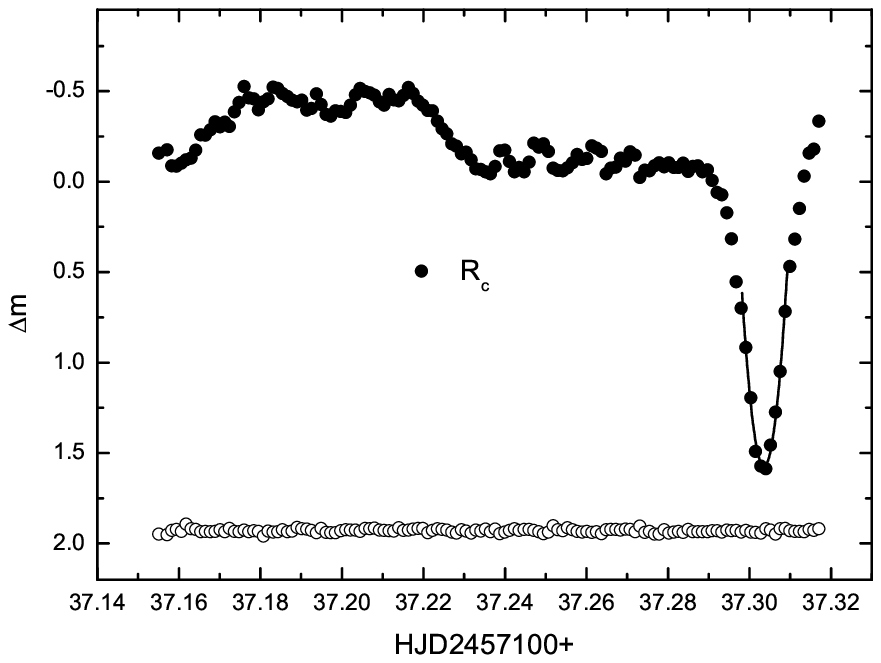}&
\includegraphics[width=8.0cm,height=7.0cm]{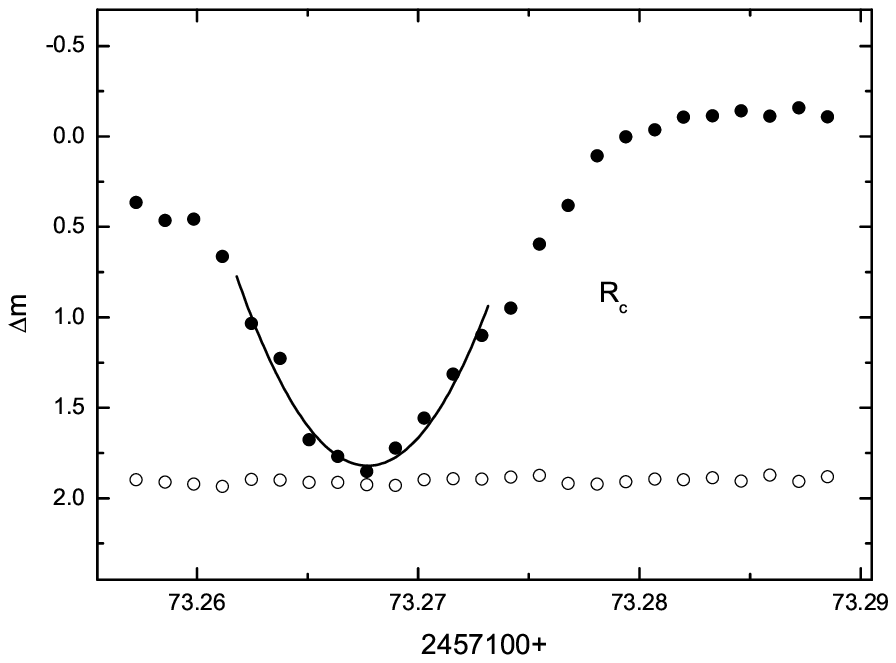}\\
\includegraphics[width=8.0cm,height=7.0cm]{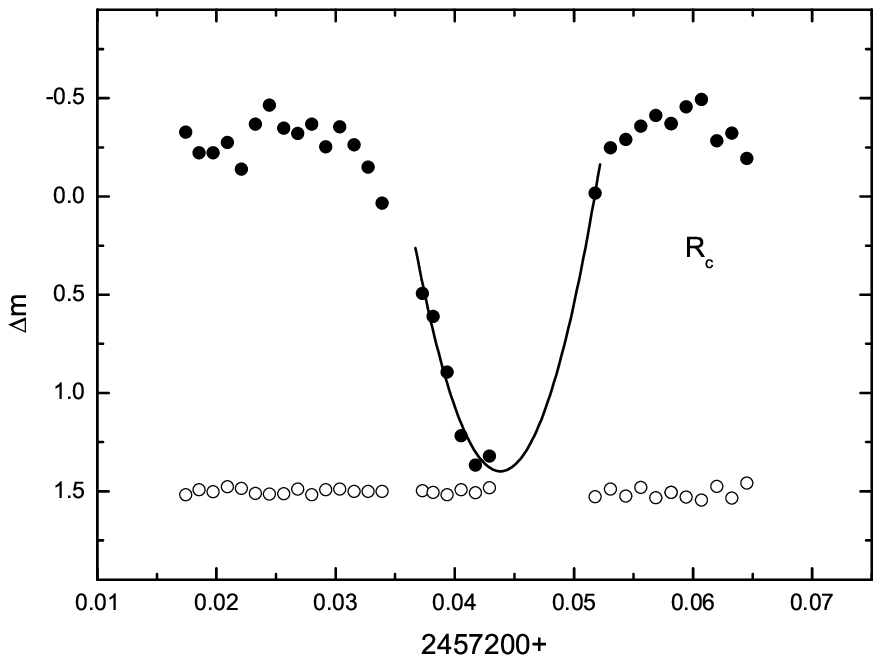}&
\includegraphics[width=8.0cm,height=7.0cm]{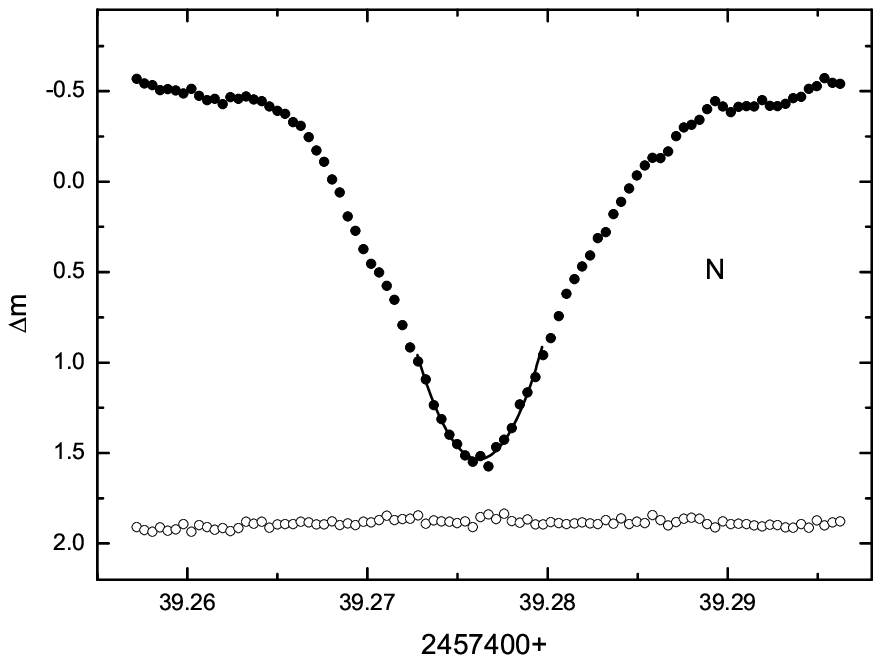}\\
\end{tabular}
\caption{The light curves of LX Ser observed using the 1.0 m Cassegrain telescope at Weihai Observatory of Shandong University. The solid and open circles represent magnitude differences between LX Ser and the comparison star and those between the comparison and the check stars, while the solid lines refer to the parabolic fits for the eclipse times. }
\end{center}
\end{figure}

\begin{table}
\scriptsize
\begin{center}
\caption{New eclipse times of LX Ser}
\begin{tabular}{cccccccccc}\hline\hline
HJD&             BJD       &Errors  & Cycle & Source&  HJD&     BJD            &Errors  & Cycle & Source\\\hline
2452777.87530 &	2452777.87605 &	0.00008 &	53555& AAVSO &  2455672.43733 &	2455672.43811 &	0.00016 &	71825& AAVSO \\
2452778.82581 &	2452778.82656 &	0.00010 &	53561& AAVSO &  2455778.42896 &	2455778.42974 &	0.00011 &	72494& AAVSO \\
2452779.77639 &	2452779.77714 &	0.00008 &	53567& AAVSO &  2456028.43535 &	2456028.43613 &	0.00010 &	74072& AAVSO \\
2452779.93490 &	2452779.93565 &	0.00012 &	53568& AAVSO &  2456088.48114 &	2456088.48192 &	0.00009 &	74451& AAVSO \\
2452780.72718 &	2452780.72793 &	0.00015 &	53573& AAVSO &  2456101.63073 &	2456101.63151 &	0.00012 &	74534& AAVSO \\
2452780.88538 &	2452780.88613 &	0.00013 &	53574& AAVSO &  2456378.88725 &	2456378.88804 &	0.00014 &	76284& AAVSO \\
2452781.83613 &	2452781.83688 &	0.00012 &	53580& AAVSO &  2456381.89715 &	2456381.89794 &	0.00019 &	76303& AAVSO \\
2452782.78677 &	2452782.78752 &	0.00012 &	53586& AAVSO &  2456383.79856 &	2456383.79935 &	0.00029 &	76315& AAVSO \\
2452782.94532 &	2452782.94607 &	0.00013 &	53587& AAVSO &  2456384.43209 &	2456384.43288 &	0.00015 &	76319& AAVSO \\
2452783.73740 &	2452783.73815 &	0.00009 &	53592& AAVSO &  2456384.59069 &	2456384.59148 &	0.00015 &	76320& AAVSO \\
2452786.74781 &	2452786.74856 &	0.00013 &	53611& AAVSO &  2456384.74908 &	2456384.74987 &	0.00016 &	76321& AAVSO \\
2452786.90609 &	2452786.90684 &	0.00009 &	53612& AAVSO &  2456385.85843 &	2456385.85922 &	0.00020 &	76328& AAVSO \\
2452787.85687 &	2452787.85762 &	0.00010 &	53618& AAVSO &  2456386.80920 &	2456386.80999 &	0.00017 &	76334& AAVSO \\
2453500.48638 &	2453500.48711 &	0.00011 &	58116& AAVSO &  2456389.50214 &	2456389.50293 &	0.00011 &	76351& AAVSO \\
2453502.54540 &	2453502.54613 &	0.00031 &	58129& AAVSO &  2456389.66136 &	2456389.66215 &	0.00021 &	76352& AAVSO \\
2453506.50728 &	2453506.50801 &	0.00033 &	58154& AAVSO &  2456403.44384 &	2456403.44463 &	0.00008 &	76439& AAVSO \\
2453514.42838 &	2453514.42911 &	0.00037 &	58204& AAVSO &  2456410.41514 &	2456410.41593 &	0.00016 &	76483& AAVSO \\
2453516.48800 &	2453516.48873 &	0.00035 &	58217& AAVSO &  2456427.68469 &	2456427.68548 &	0.00020 &	76592& AAVSO \\
2453519.49811 &	2453519.49884 &	0.00042 &	58236& AAVSO &  2456782.41510 &	2456782.41588 &	0.00006 &	78831& AAVSO \\
2453521.39970 &	2453521.40043 &	0.00039 &	58248& AAVSO &  2456792.39590 &	2456792.39668 &	0.00020 &	78894& AAVSO \\
2453521.55769 &	2453521.55842 &	0.00040 &	58249& AAVSO &  2456798.41636 &	2456798.41714 &	0.00006 &	78932& AAVSO \\
2453541.52033 &	2453541.52106 &	0.00012 &	58375& AAVSO &  2457091.67550 &	2457091.67627 &	0.00007 &	80783& AAVSO \\
2454316.41428 &	2454316.41502 &	0.00008 &	63266& AAVSO &  2457094.68559 &	2457094.68636 &	0.00021 &	80802& AAVSO \\
2454580.52079 &	2454580.52153 &	0.00034 &	64933& AAVSO &  2457097.69575 &	2457097.69652 &	0.00005 &	80821& AAVSO \\
2454628.52582 &	2454628.52656 &	0.00013 &	65236& AAVSO &  2457134.45164 &	2457134.45241 &	0.00004 &	81053& AAVSO \\
2454976.44292 &	2454976.44368 &	0.00014 &	67432& AAVSO &  2457137.30350 &	2457137.30427 &	0.00007 &	81071& 1m    \\
2454994.50413 &	2454994.50489 &	0.00006 &	67546& AAVSO &  2457159.48411 &	2457159.48488 &	0.00005 &	81211& AAVSO \\
2455001.47512 &	2455001.47588 &	0.00007 &	67590& AAVSO &  2457163.44506 &	2457163.44583 &	0.00013 &	81236& AAVSO \\
2455037.43966 &	2455037.44042 &	0.00017 &	67817& AAVSO &  2457173.26766 &	2457173.26843 &	0.00015 &	81298& 1m    \\
2455662.45633 &	2455662.45711 &	0.00016 &	71762& AAVSO &  2457200.04314 &	2457200.04391 &	0.00049 &	81467& 1m    \\
2455663.40660 &	2455663.40738 &	0.00022 &	71768& AAVSO &  2457439.27618 &	2457439.27695 &	0.00004 &	82977& 1m    \\

\hline
\end{tabular}
\end{center}
\end{table}

\begin{figure}
\begin{center}
\begin{tabular}{c@{\hspace{0.3pc}}c}
\includegraphics[width=4.0cm]{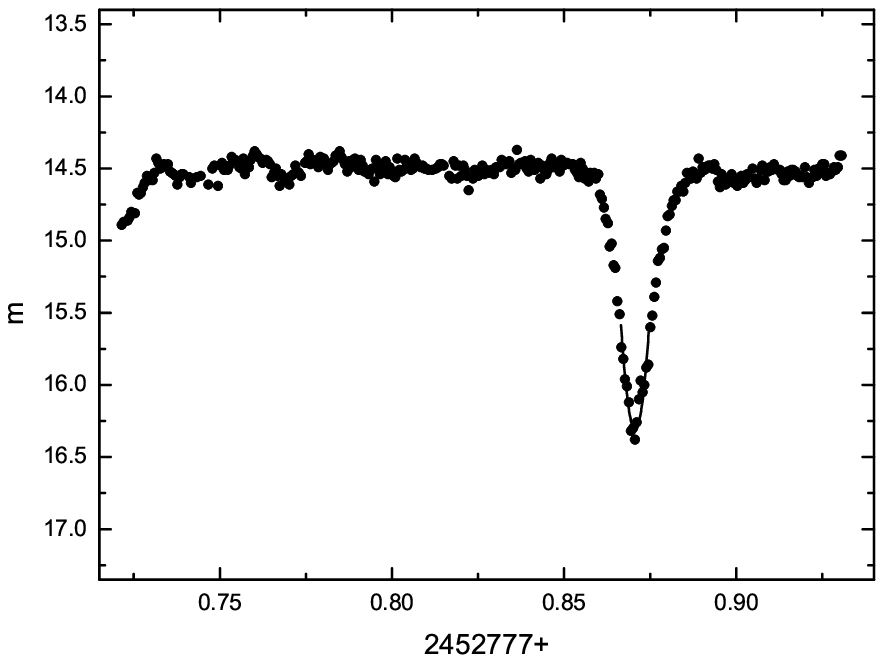}
\includegraphics[width=4.0cm]{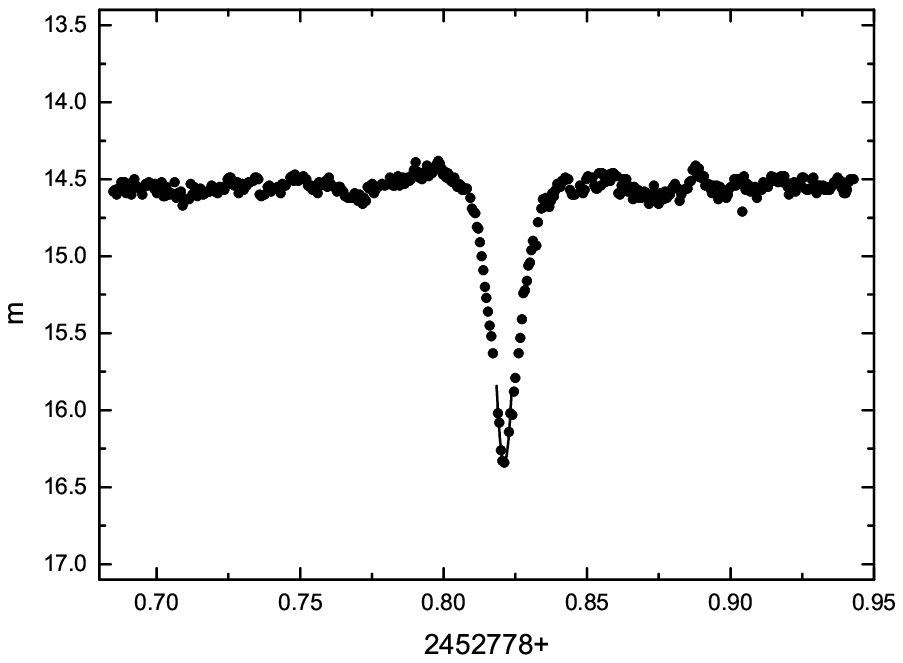}&
\includegraphics[width=4.0cm]{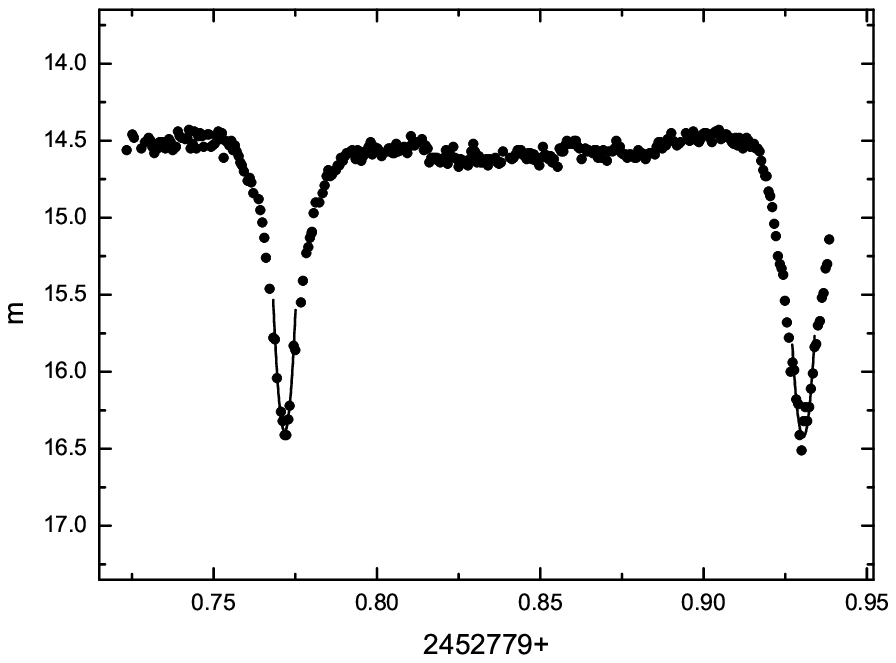}
\includegraphics[width=4.0cm]{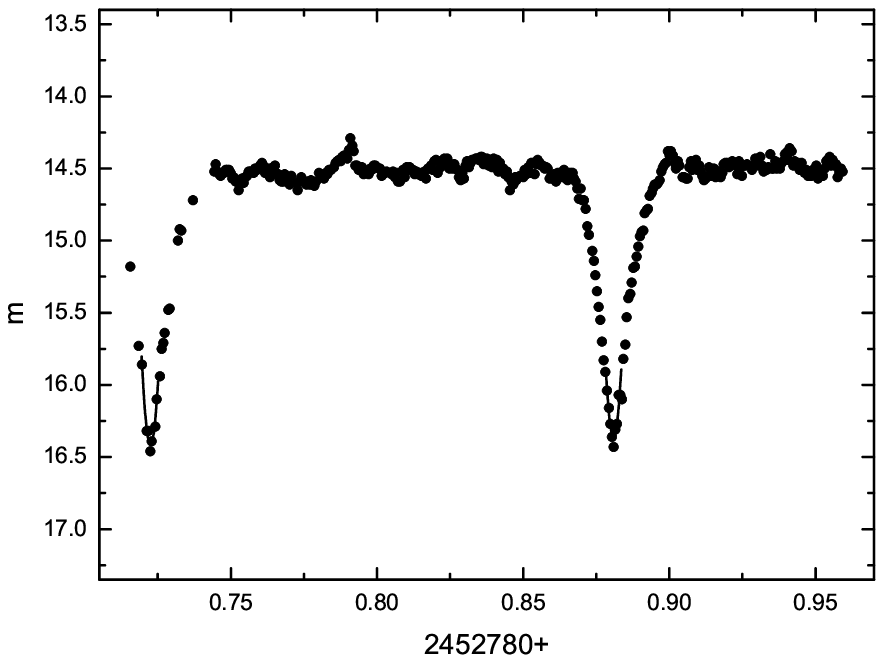}\\
\includegraphics[width=4.0cm]{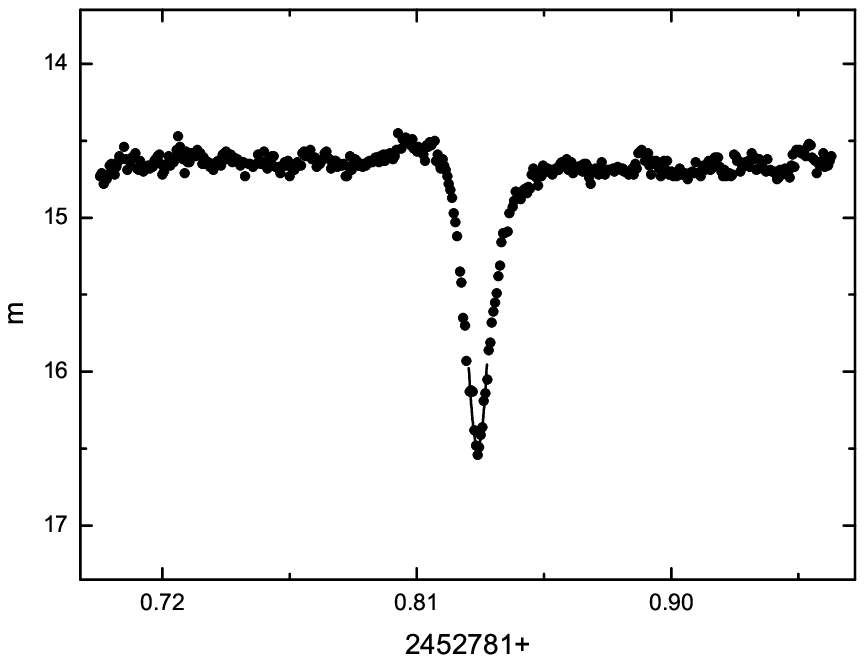}
\includegraphics[width=4.0cm]{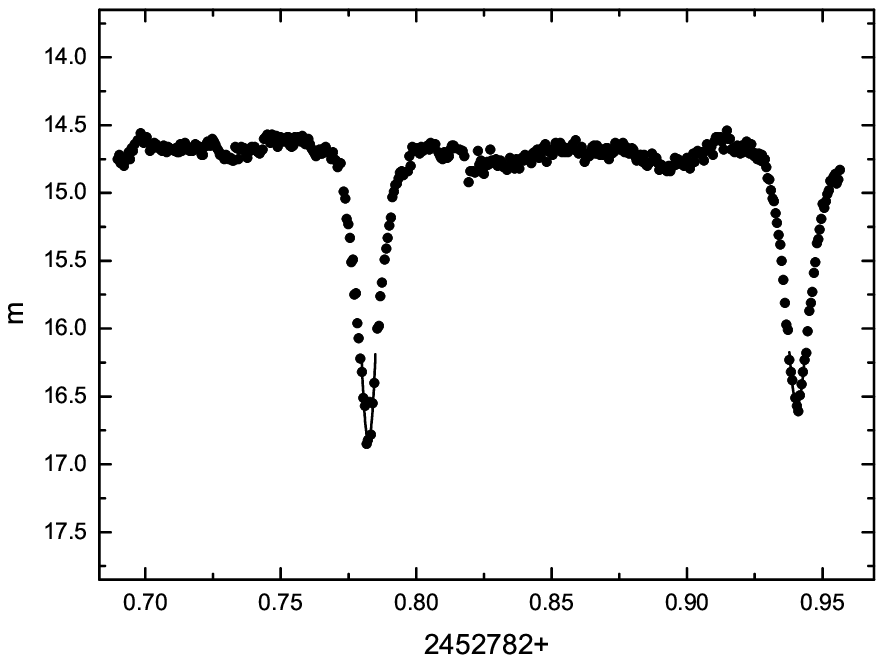}&
\includegraphics[width=4.0cm]{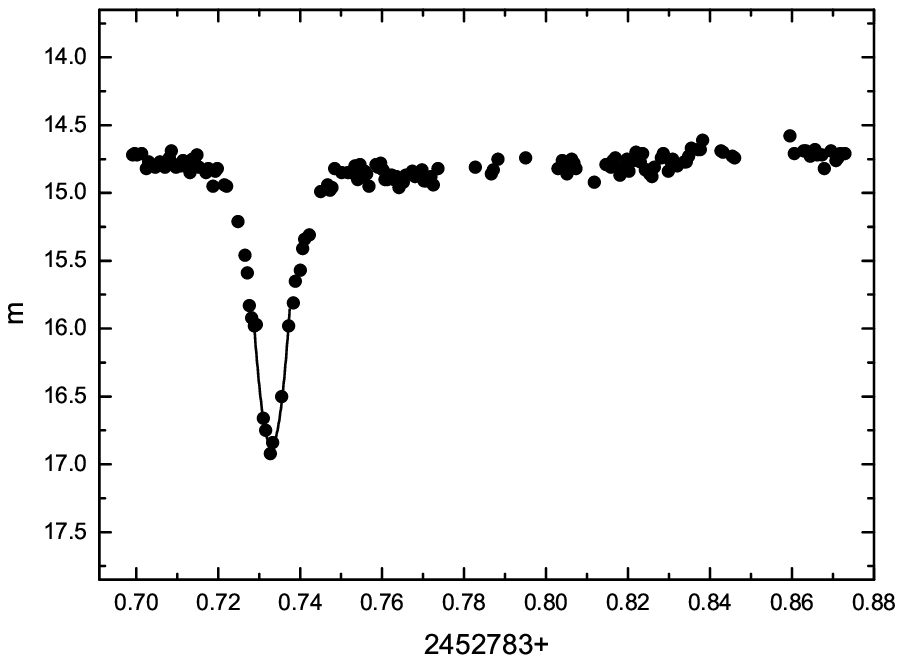}
\includegraphics[width=4.0cm]{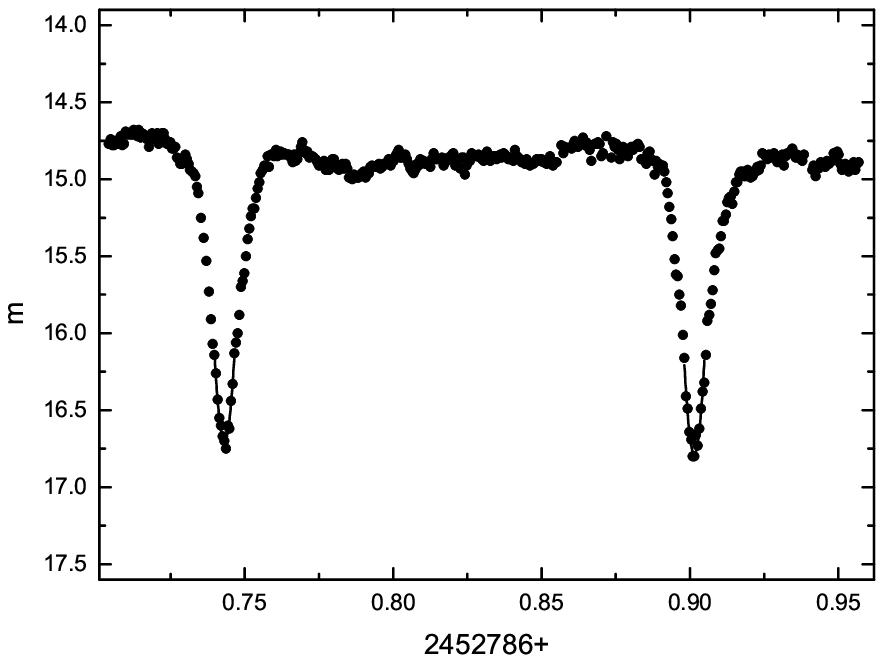}\\
\includegraphics[width=4.0cm]{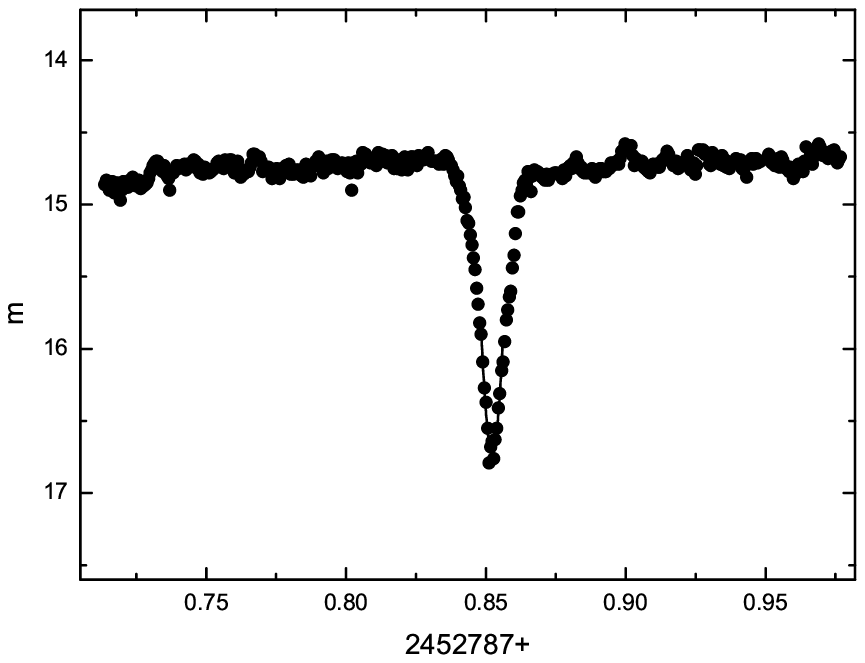}
\includegraphics[width=4.0cm]{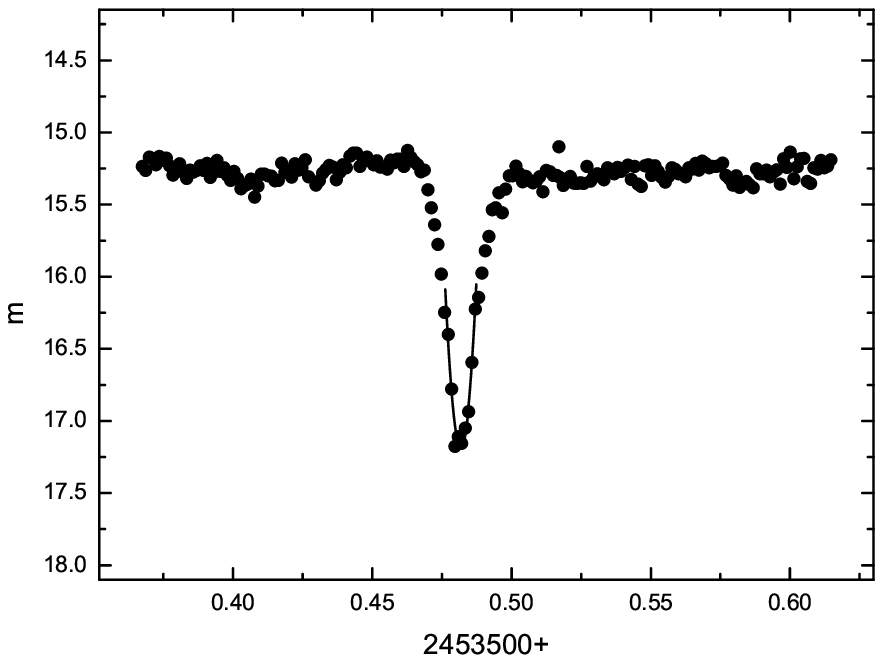}&
\includegraphics[width=4.0cm]{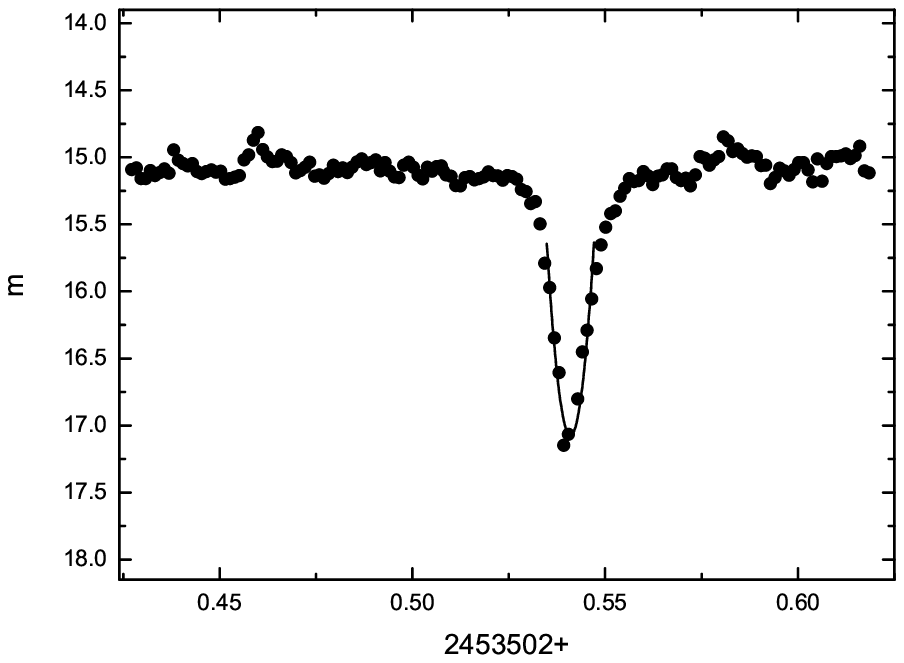}
\includegraphics[width=4.0cm]{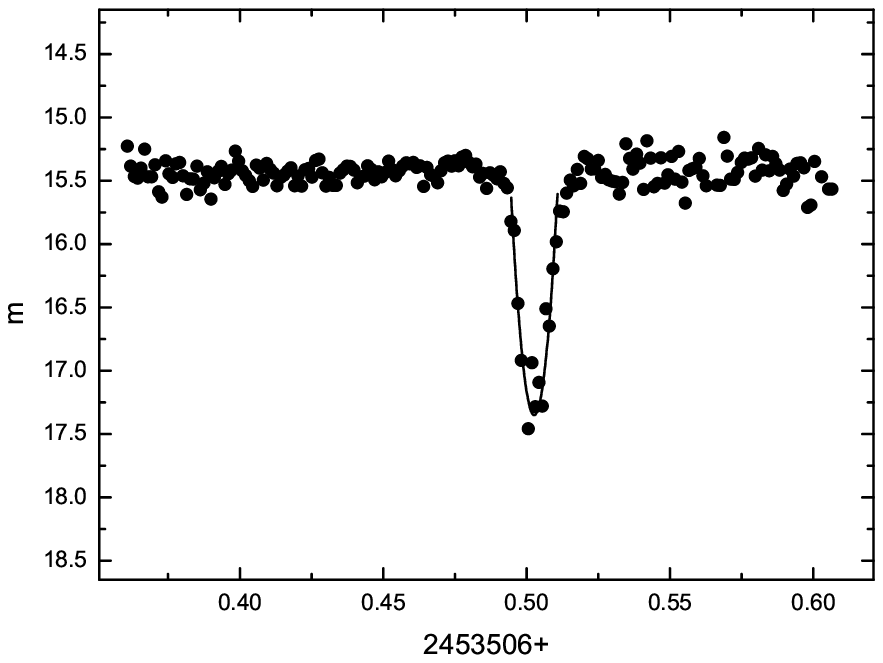}\\
\includegraphics[width=4.0cm]{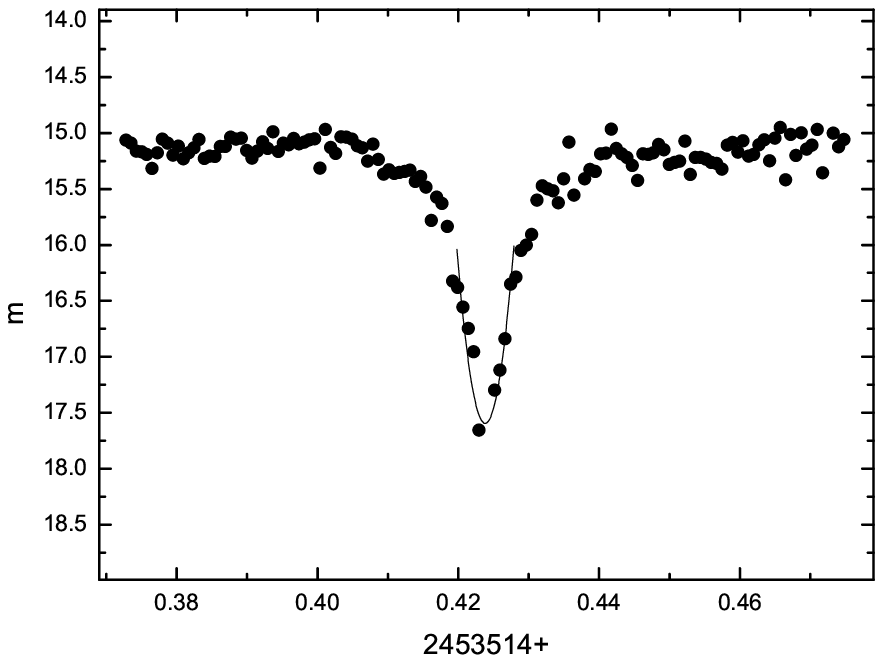}
\includegraphics[width=4.0cm]{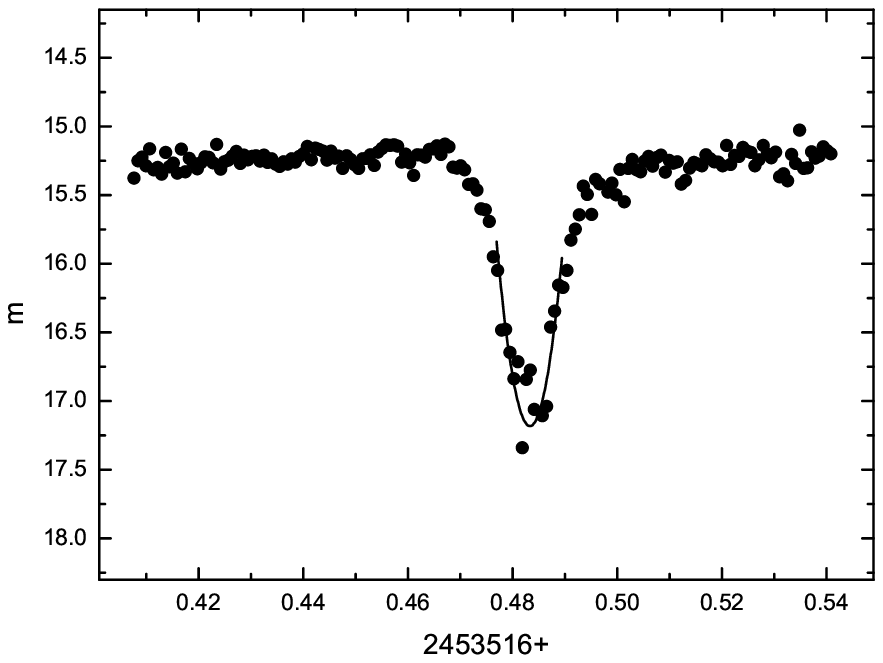}&
\includegraphics[width=4.0cm]{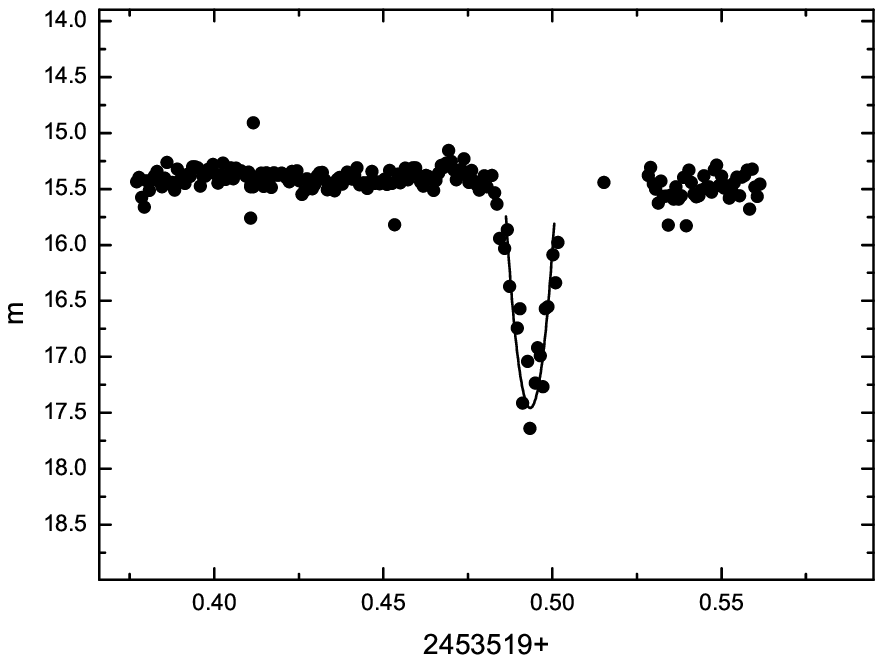}
\includegraphics[width=4.0cm]{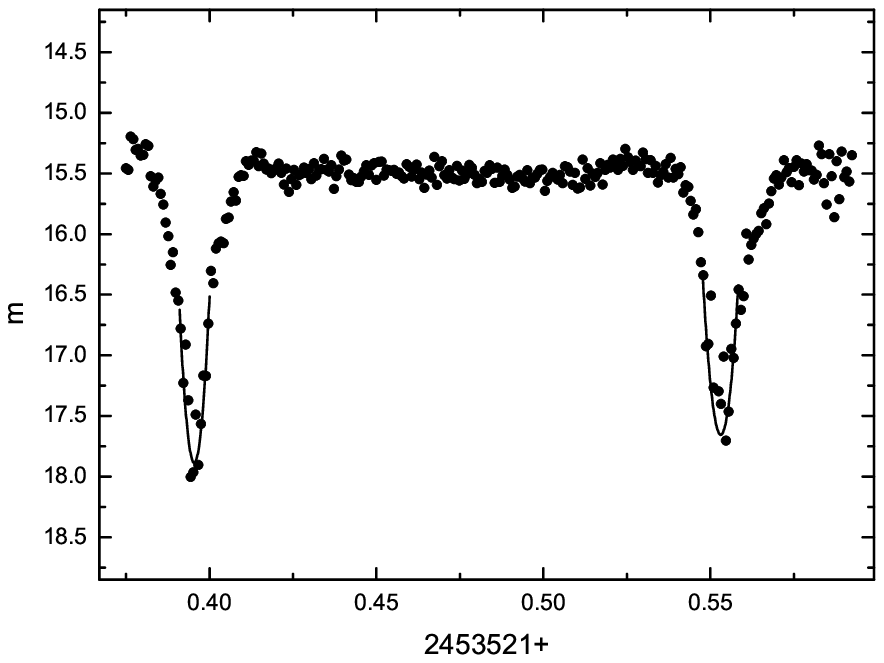}\\
\includegraphics[width=4.0cm]{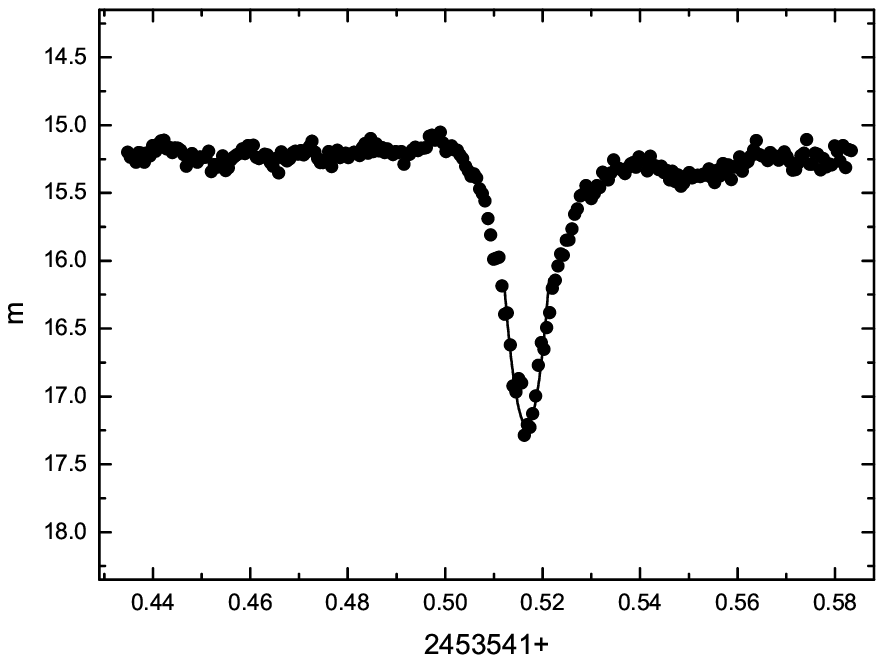}
\includegraphics[width=4.0cm]{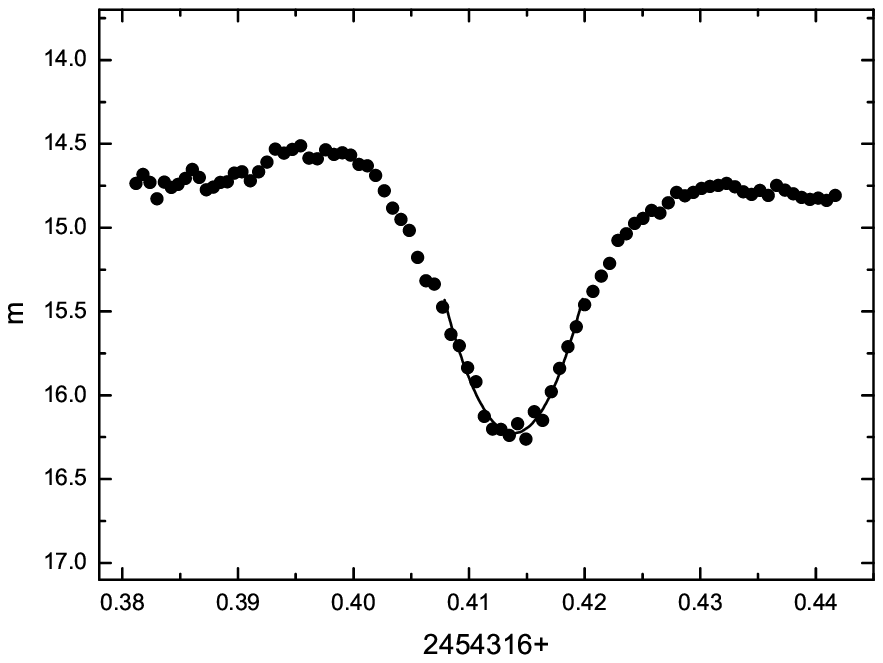}&
\includegraphics[width=4.0cm]{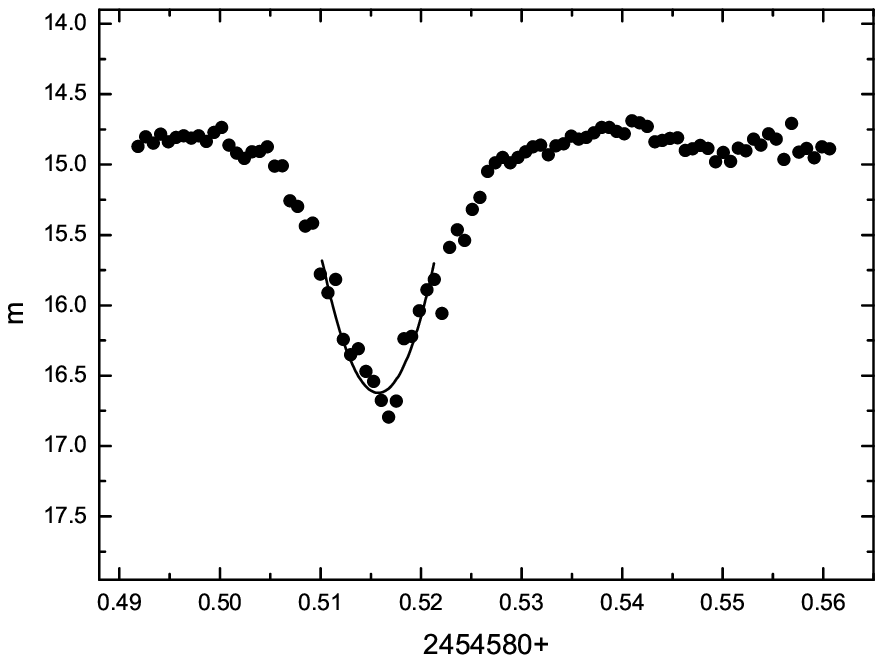}
\includegraphics[width=4.0cm]{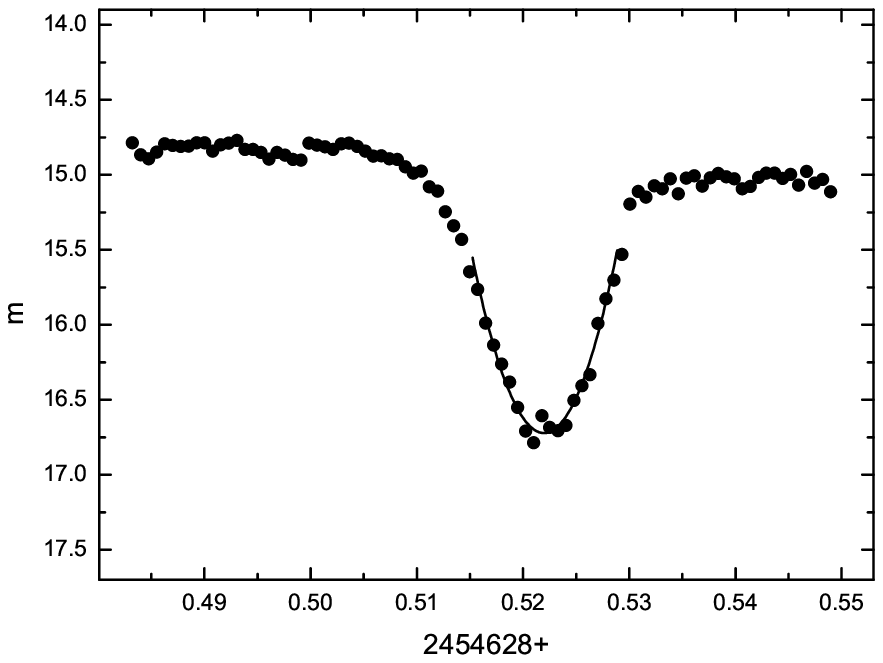}\\
\includegraphics[width=4.0cm]{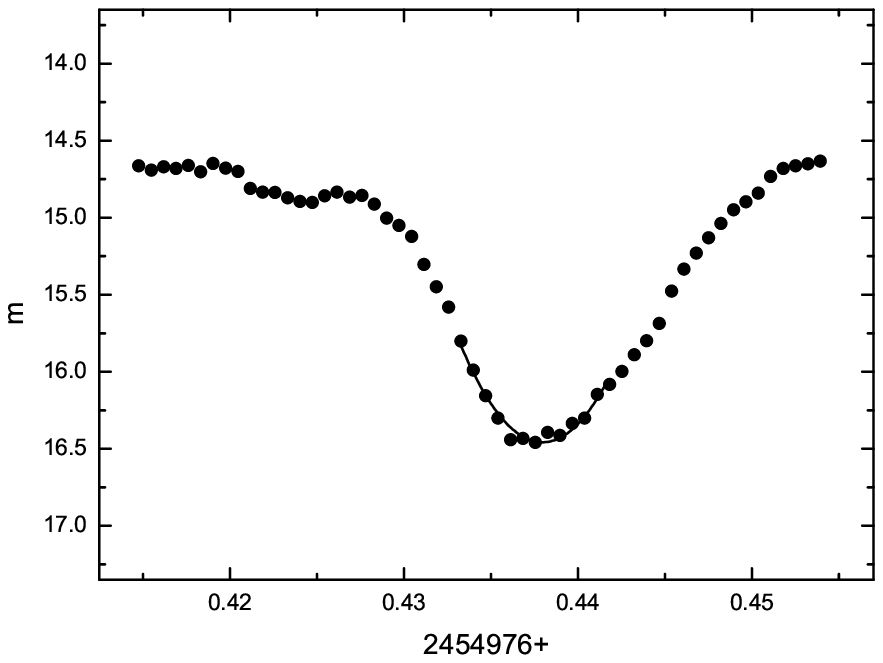}
\includegraphics[width=4.0cm]{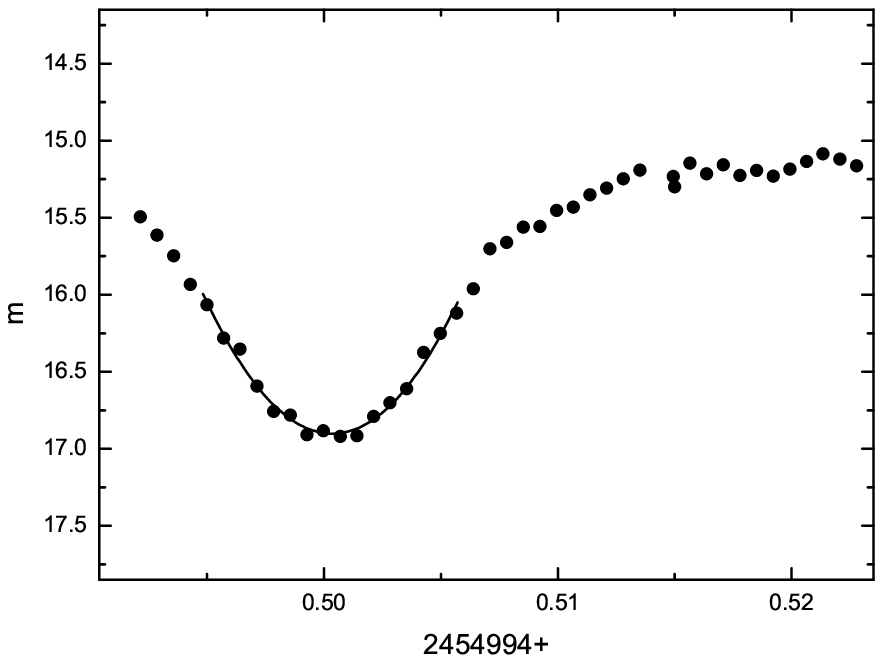}&
\includegraphics[width=4.0cm]{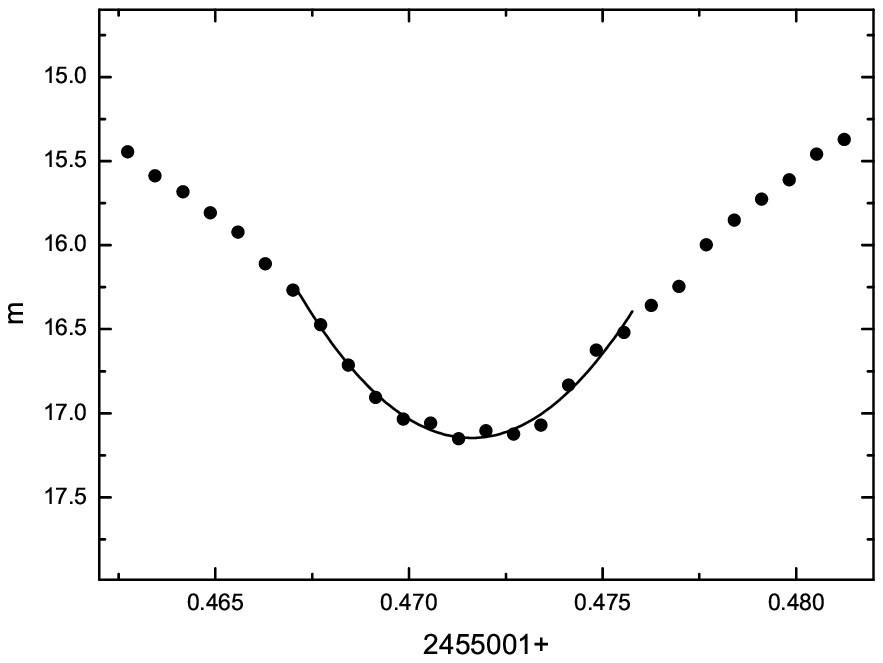}
\includegraphics[width=4.0cm]{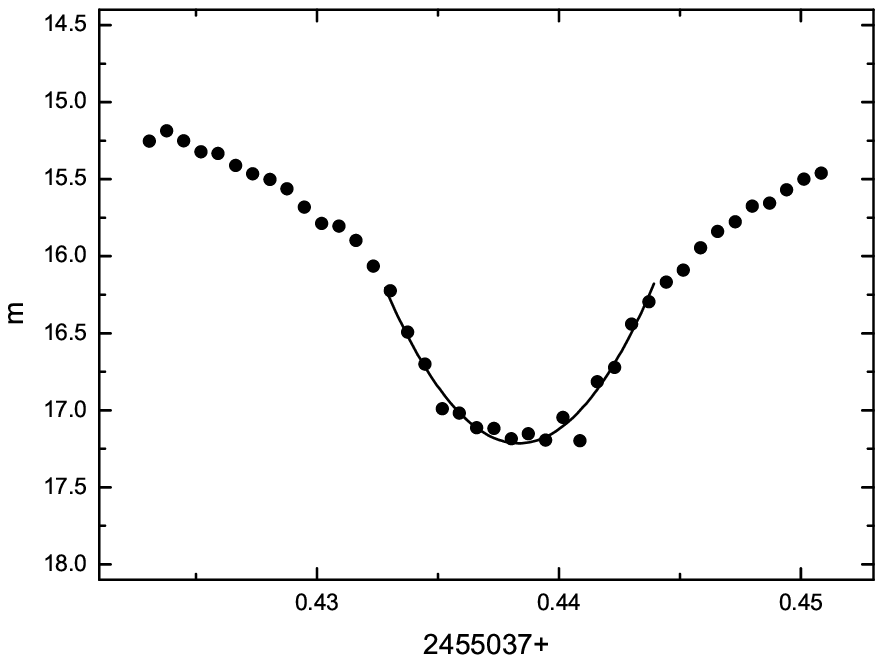}\\
\end{tabular}
\end{center}
\caption{The light curves of LX Ser using the AAVSO data. The solid circles represent the visual magnitude of LX Ser, while the solid lines refer to the parabolic fits for the eclipse times. }
\end{figure}

\addtocounter{figure}{-1}
\begin{figure}
\begin{center}
\begin{tabular}{c@{\hspace{0.3pc}}c}
\includegraphics[width=4.0cm]{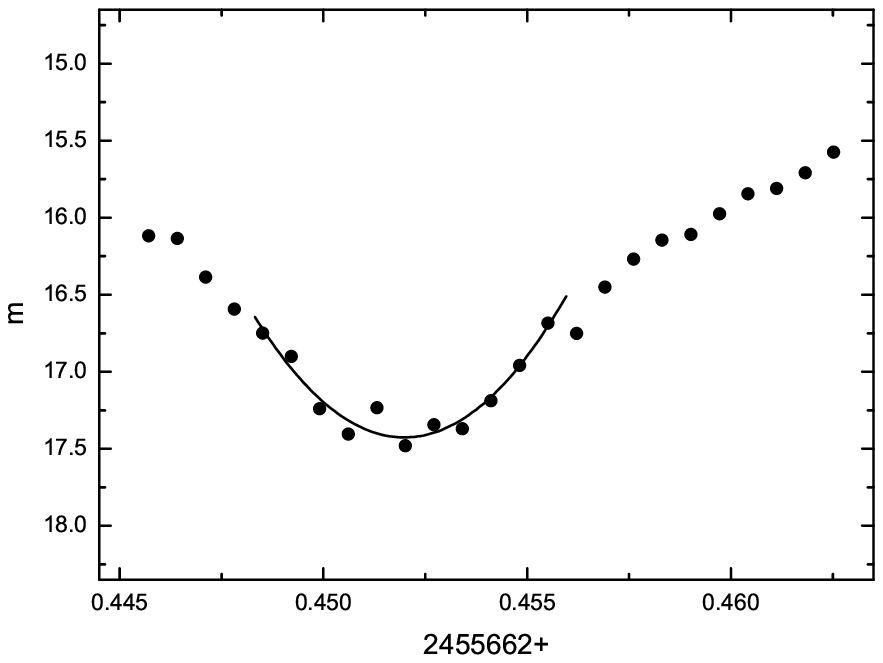}
\includegraphics[width=4.0cm]{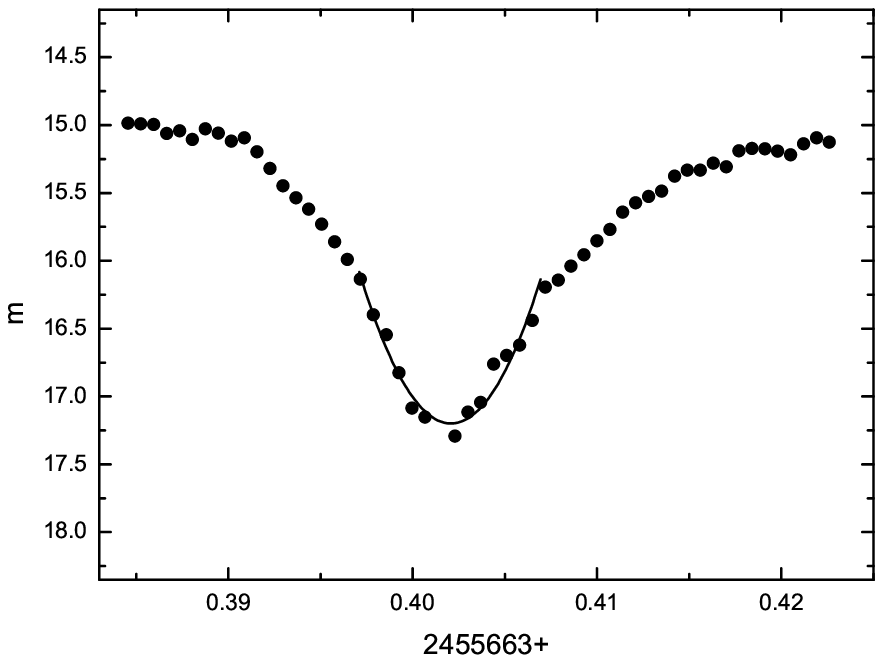}
\includegraphics[width=4.0cm]{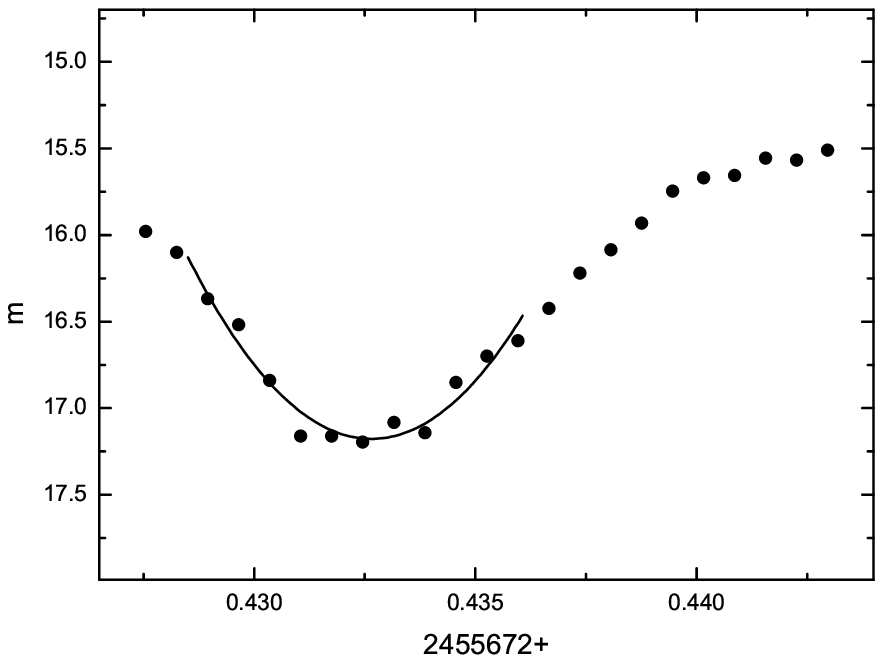}&
\includegraphics[width=4.0cm]{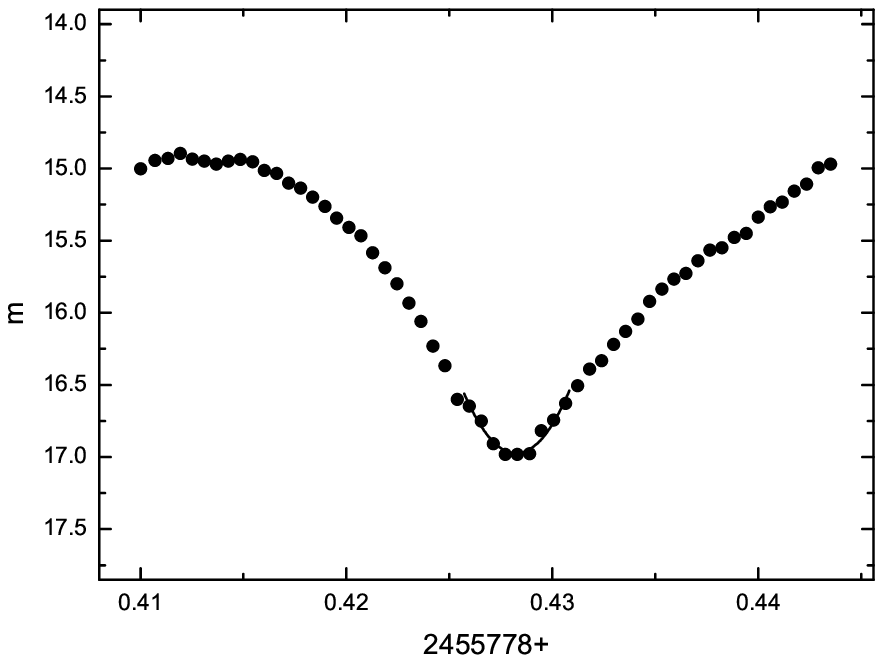}\\
\includegraphics[width=4.0cm]{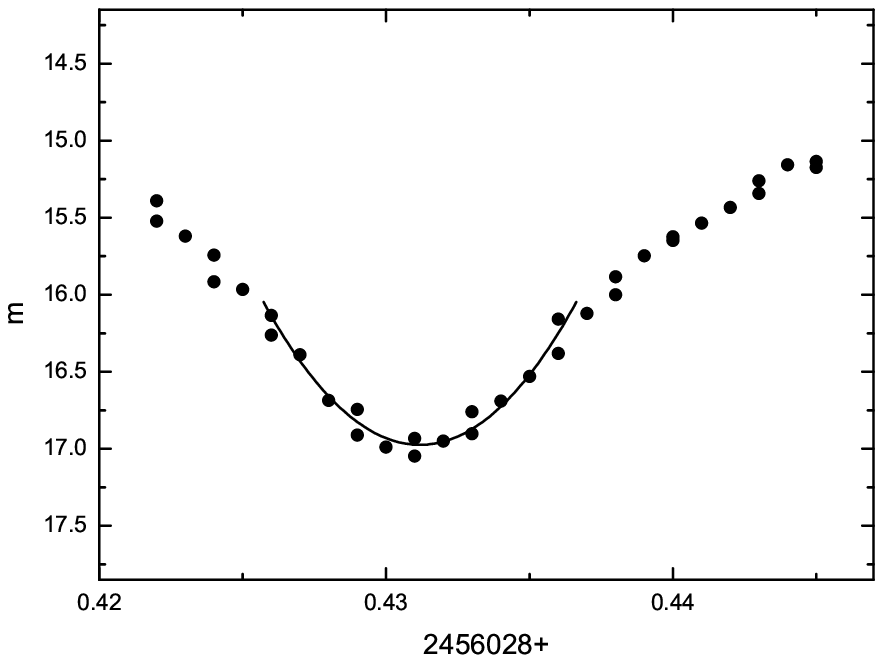}
\includegraphics[width=4.0cm]{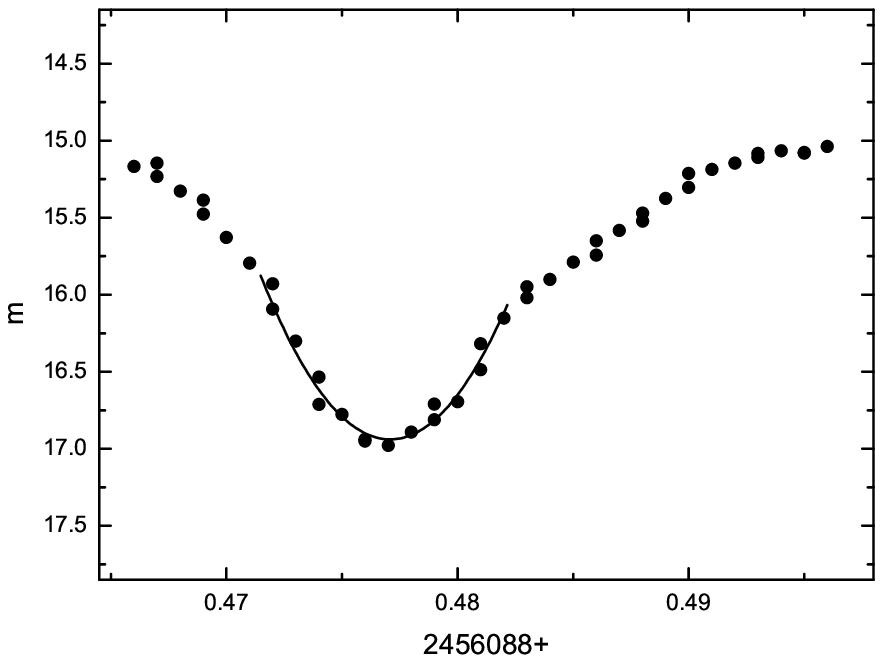}
\includegraphics[width=4.0cm]{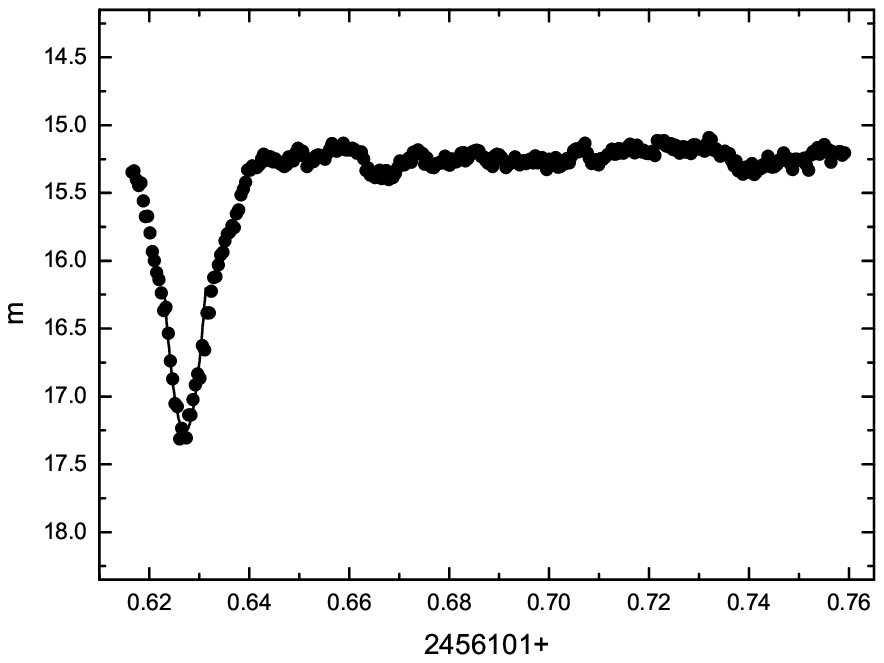}&
\includegraphics[width=4.0cm]{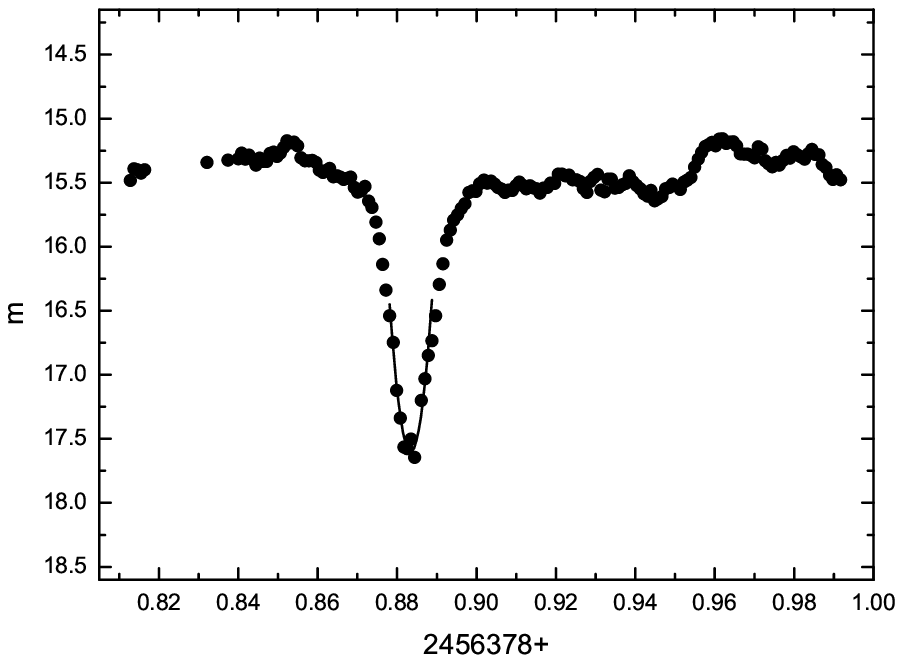}\\
\includegraphics[width=4.0cm]{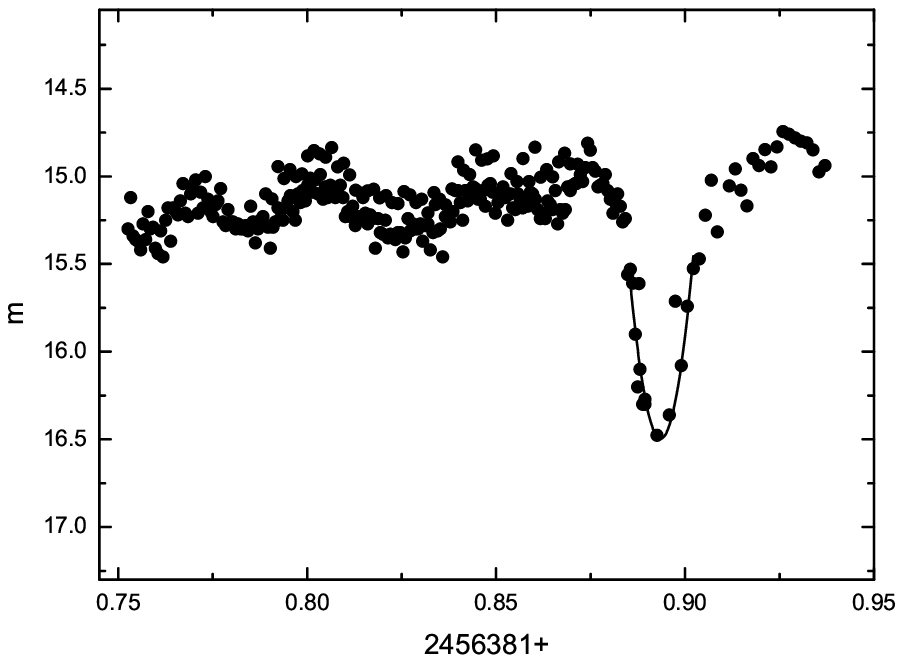}
\includegraphics[width=4.0cm]{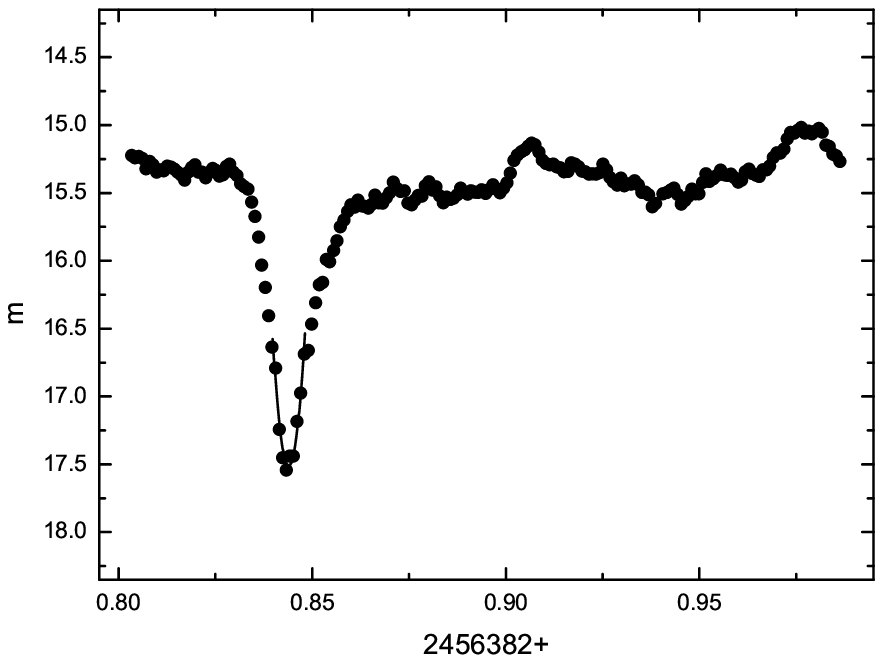}
\includegraphics[width=4.0cm]{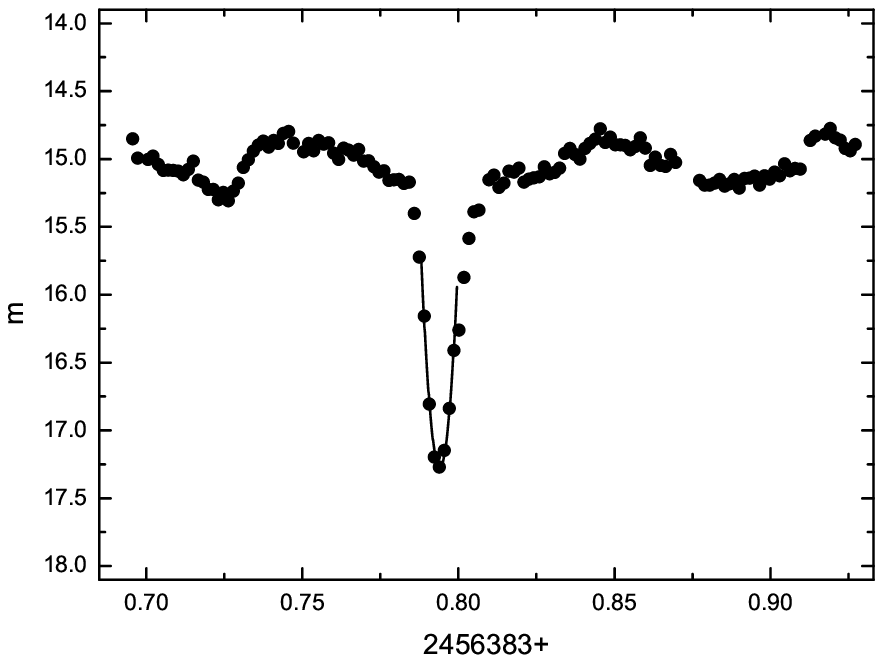}&
\includegraphics[width=4.0cm]{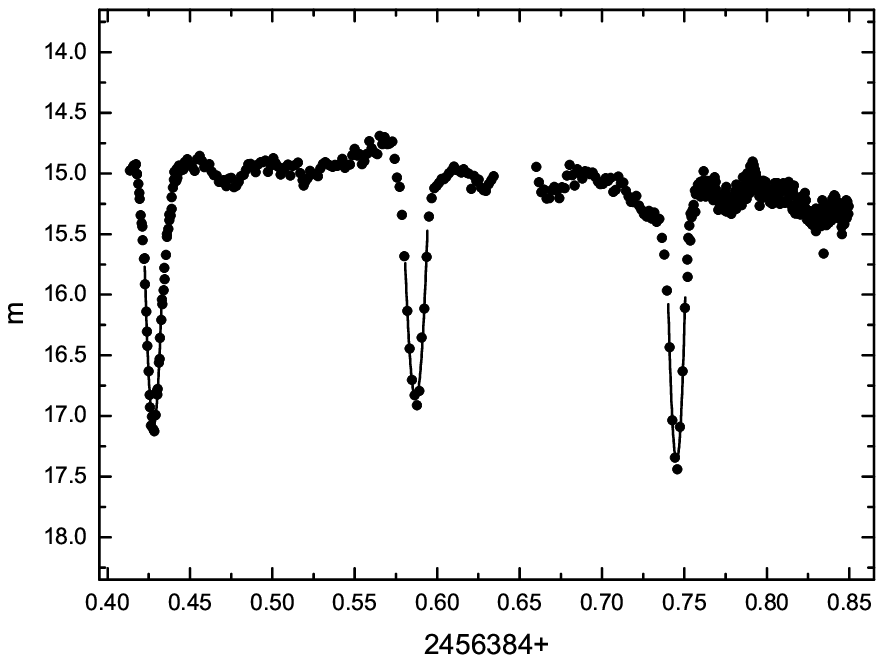}\\
\includegraphics[width=4.0cm]{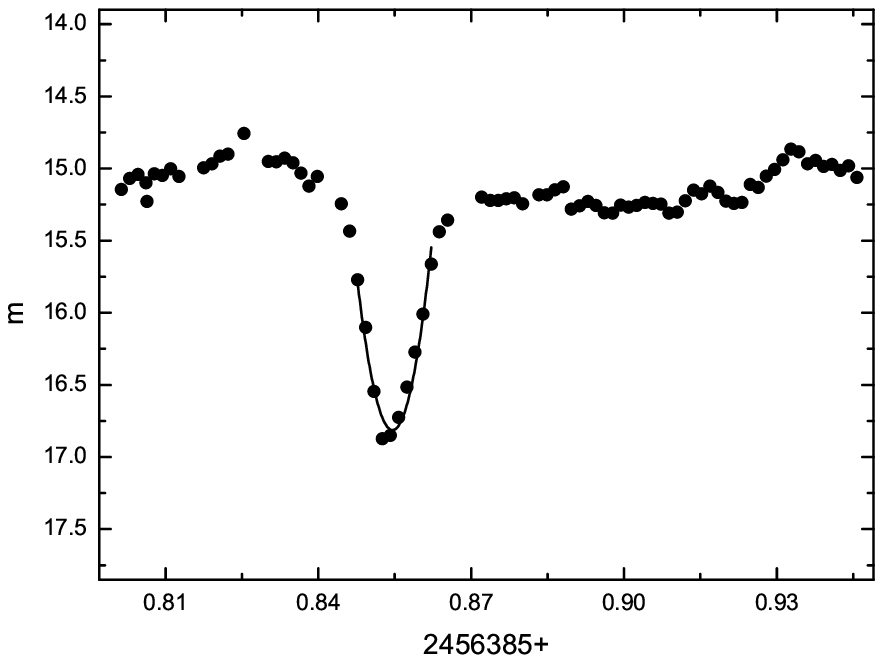}
\includegraphics[width=4.0cm]{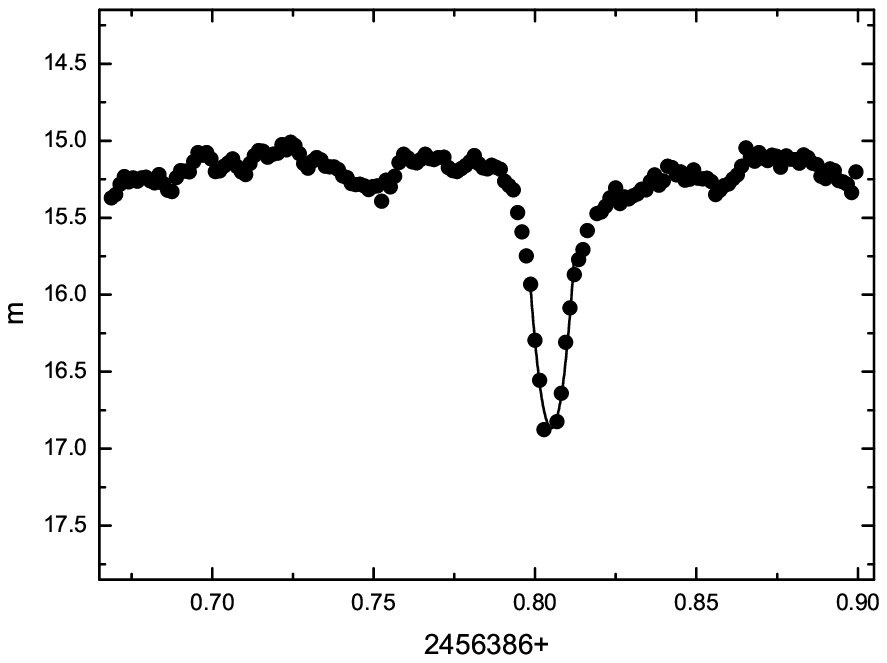}
\includegraphics[width=4.0cm]{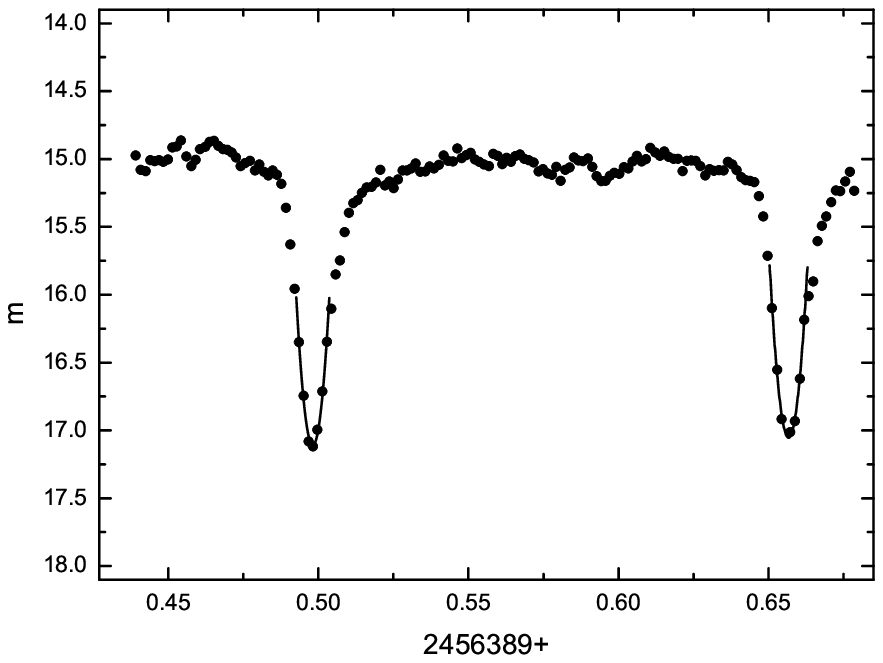}&
\includegraphics[width=4.0cm]{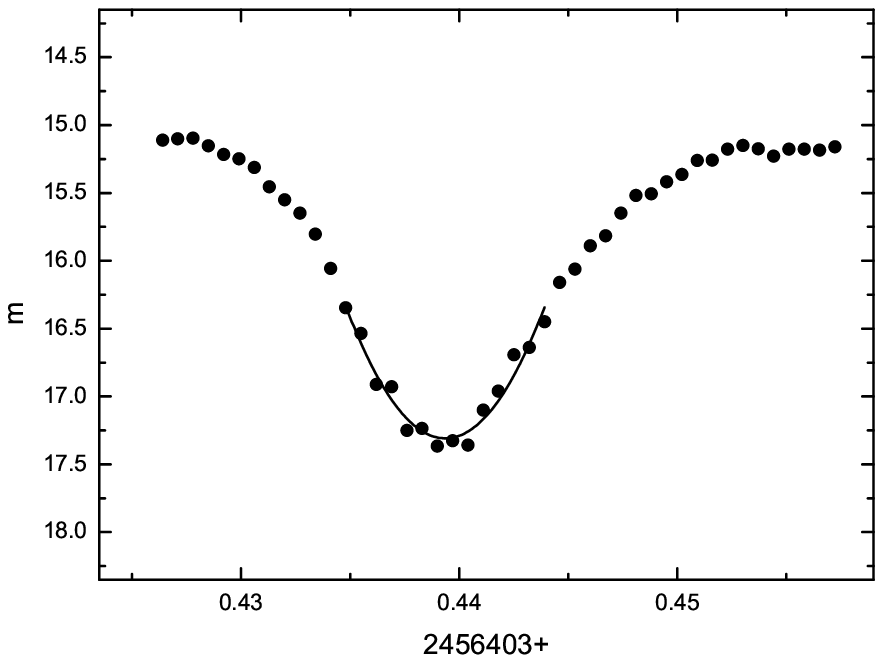}\\
\includegraphics[width=4.0cm]{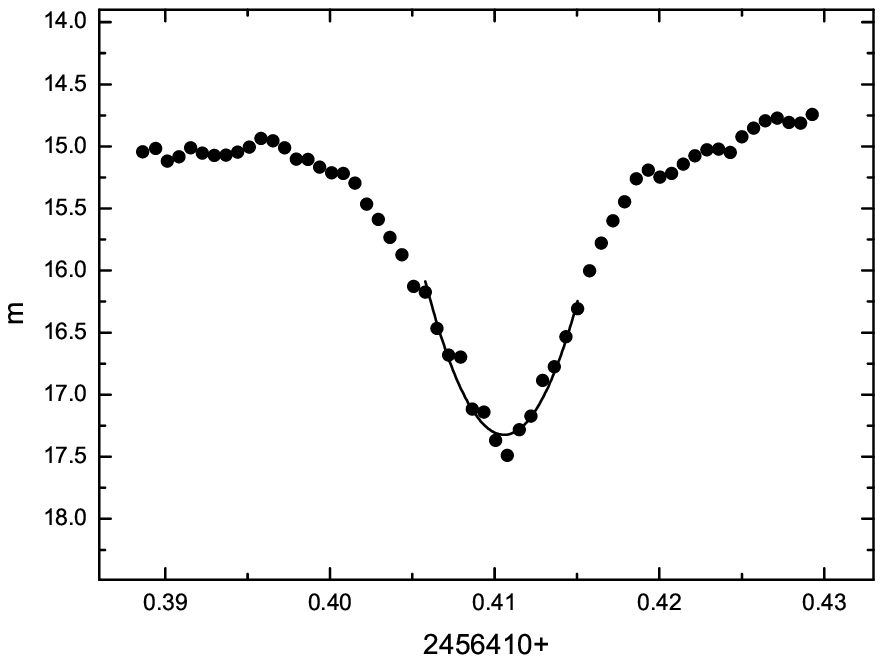}
\includegraphics[width=4.0cm]{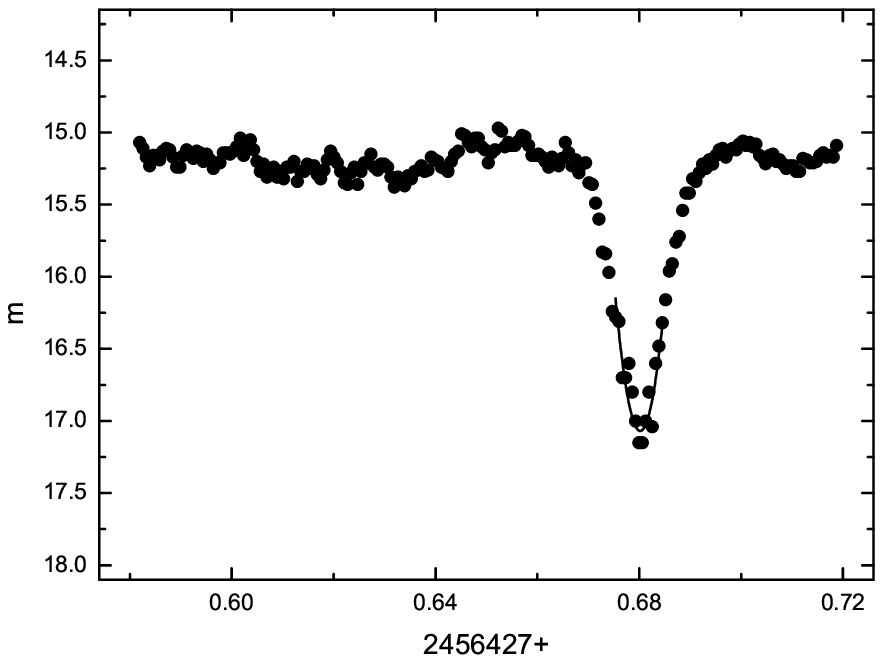}
\includegraphics[width=4.0cm]{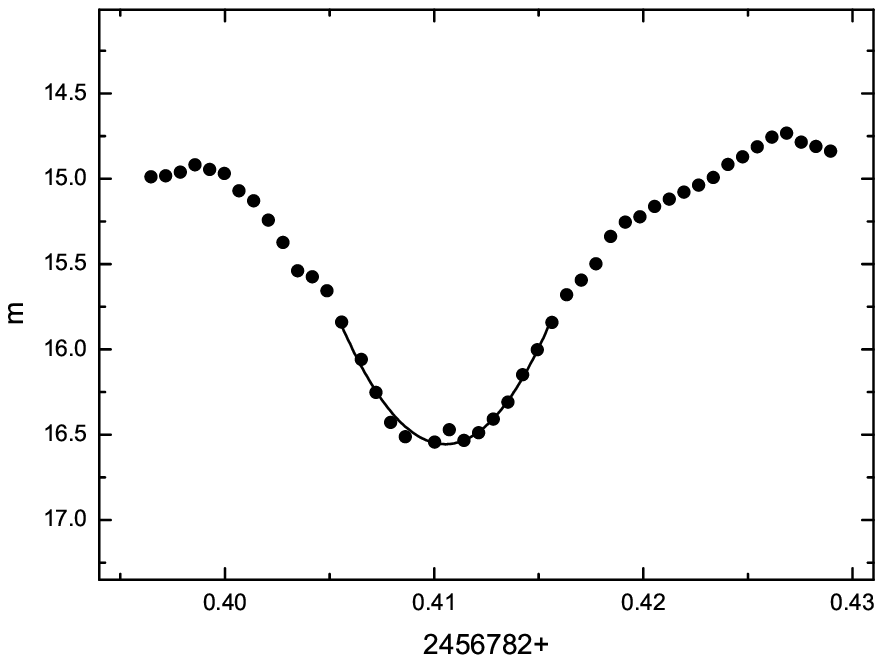}&
\includegraphics[width=4.0cm]{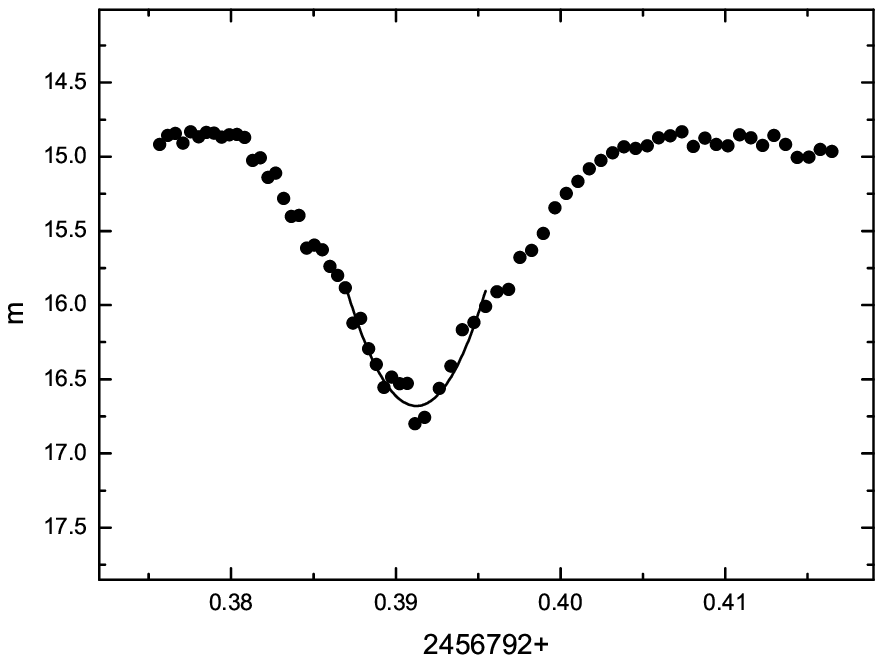}\\
\includegraphics[width=4.0cm]{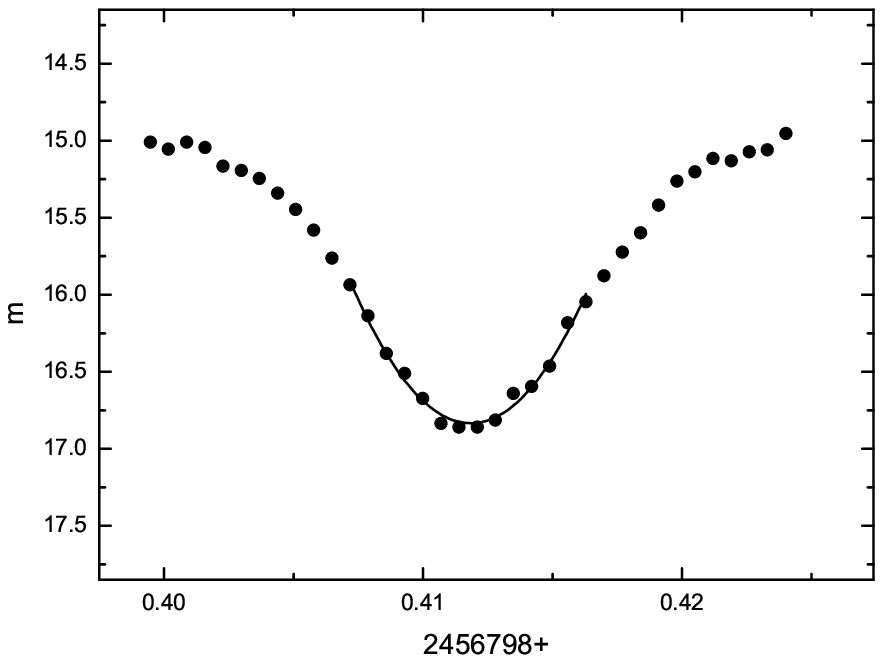}
\includegraphics[width=4.0cm]{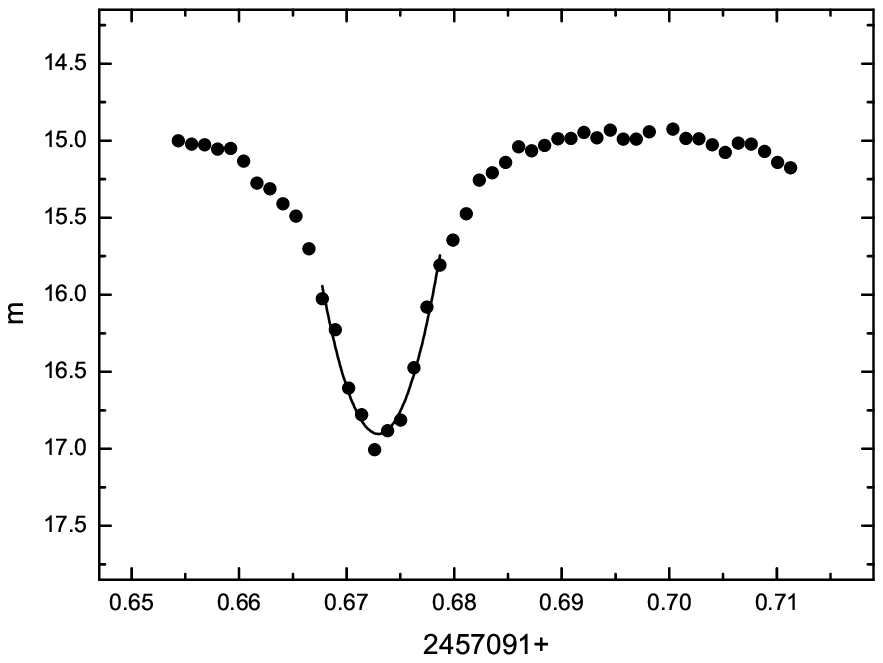}
\includegraphics[width=4.0cm]{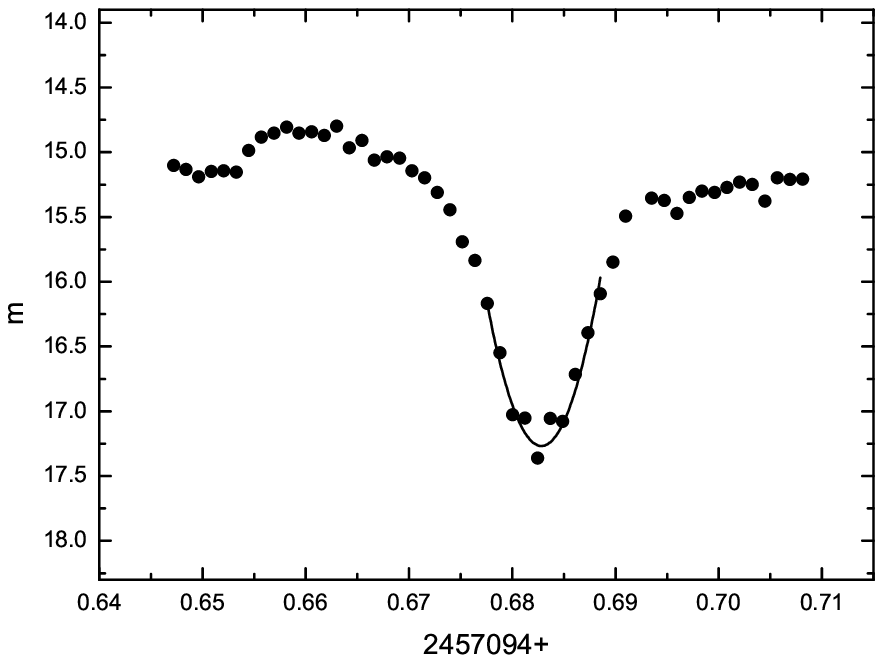}&
\includegraphics[width=4.0cm]{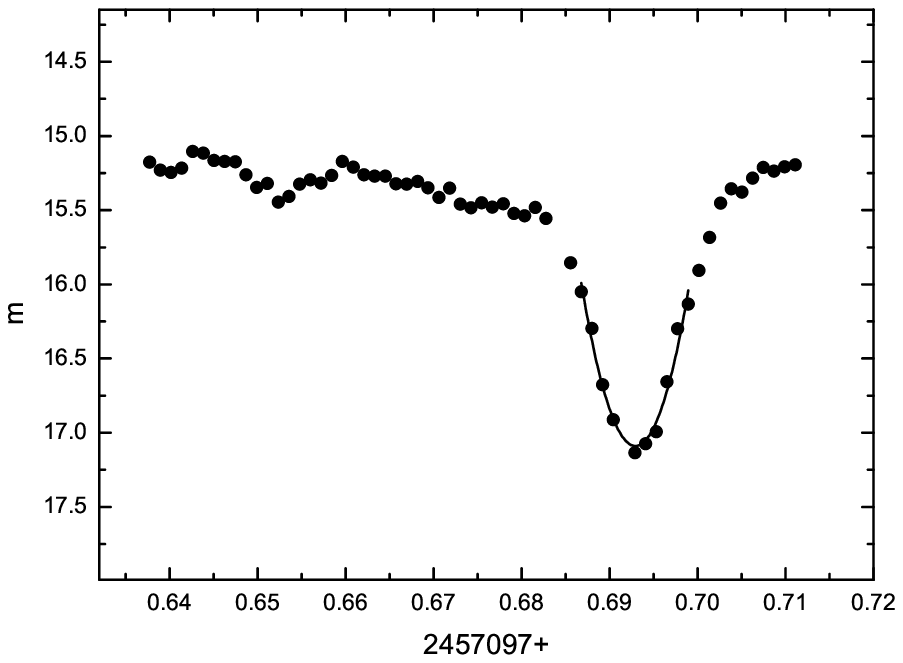}\\
\includegraphics[width=4.0cm]{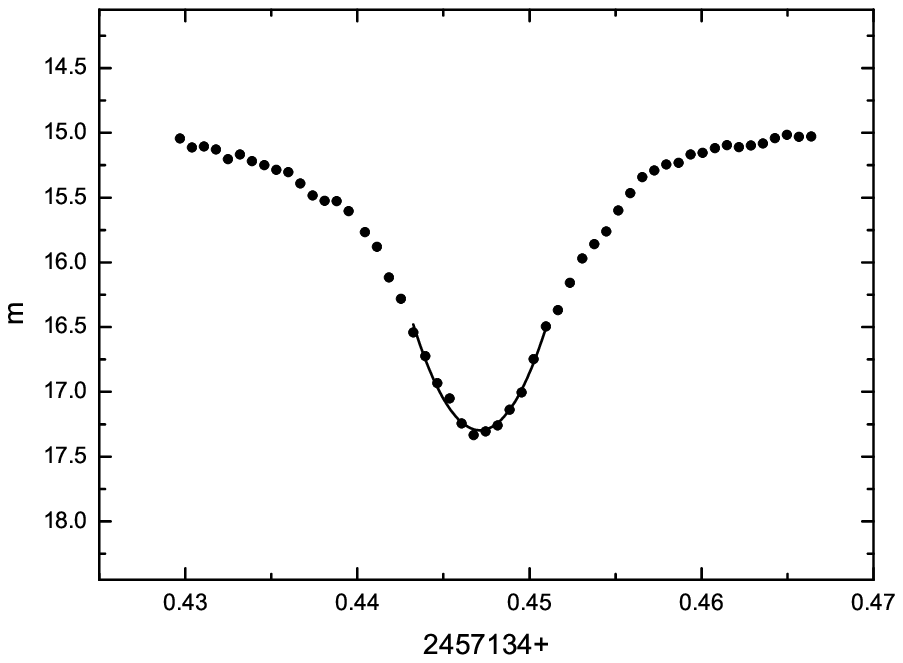}
\includegraphics[width=4.0cm]{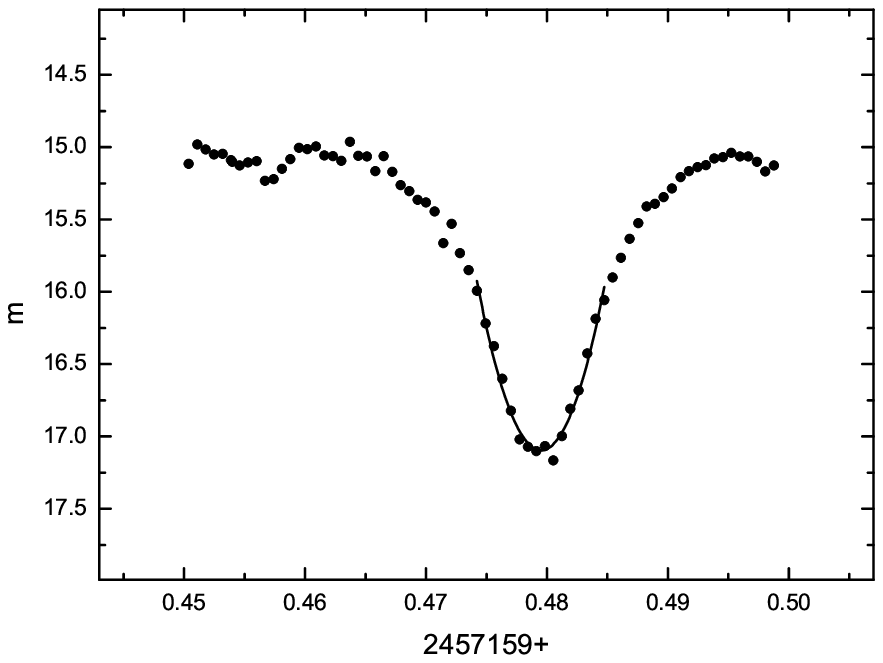}
\includegraphics[width=4.0cm]{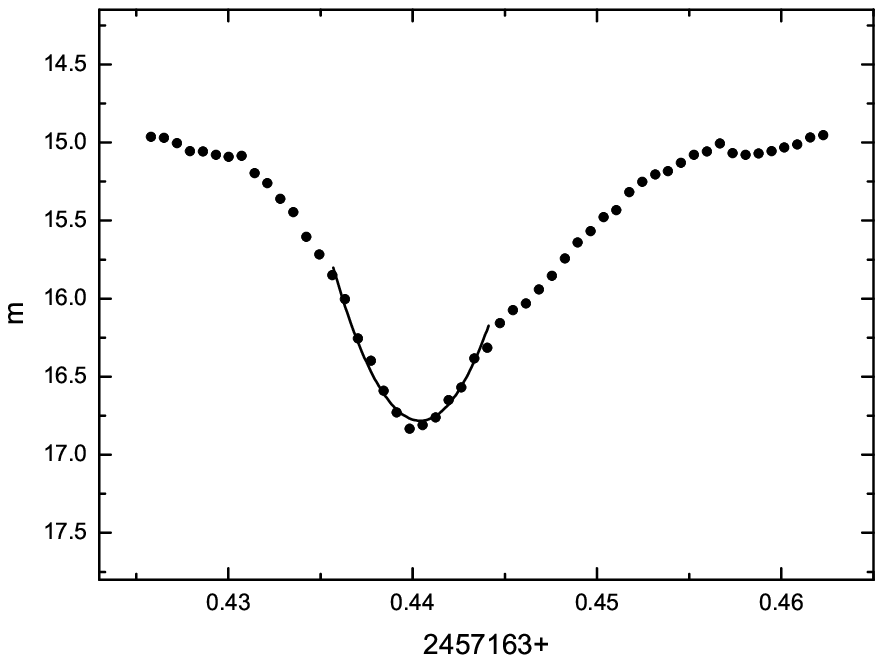}&\\
\end{tabular}
\caption{--- continued }
\end{center}
\end{figure}

\section{Investigation of the $O-C$ diagram}
Since the discovery, LX Ser has been continuously monitored. Many highly precise eclipse times were determined. All the eclipse times collected from literatures including our 62 new determined ones are listed in Table 3. The eclipse times collected form literatures are checked again since some data were derived a long time ago and obtained by different observatories or filters. Those having obviously erring are abandoned. Some of the eclipse times were not given errors in the literatures. The errors of the visual data are assumed to be 0.00100, while that of the photoelectric and CCD data are 0.00010. We constructed the $O-C$ diagram using the following linear ephemeris taken from O-C Gateway\footnote{http://var2.astro.cz/ocgate/} (the initial epoch was converted to BJD),
\begin{eqnarray}
Min.I = BJD2444293.02457 + 0^d.1584325E.
\end{eqnarray}
The $O-C$ values are listed in Table 3 and the corresponding $O-C$ curve is plotted in the upper panel of Figure 3. A linear function was used to fit the $O-C$ values. By using the weighted least-squares method, we determined the following equation
\begin{eqnarray}
Min.I = BJD2444293.024192(\pm0.000083) + 0^d.158432491(\pm0.000000002)\times E .
\end{eqnarray}
The variance is $\sigma_1=5.14\times10^{-7}$ and the fitting curve is displayed in Figure 3. During the fitting, the weight for the visual data is 1, while that for the photoelectric and CCD data is 10.
\begin{figure}
\begin{center}
\includegraphics[angle=0,scale=0.7]{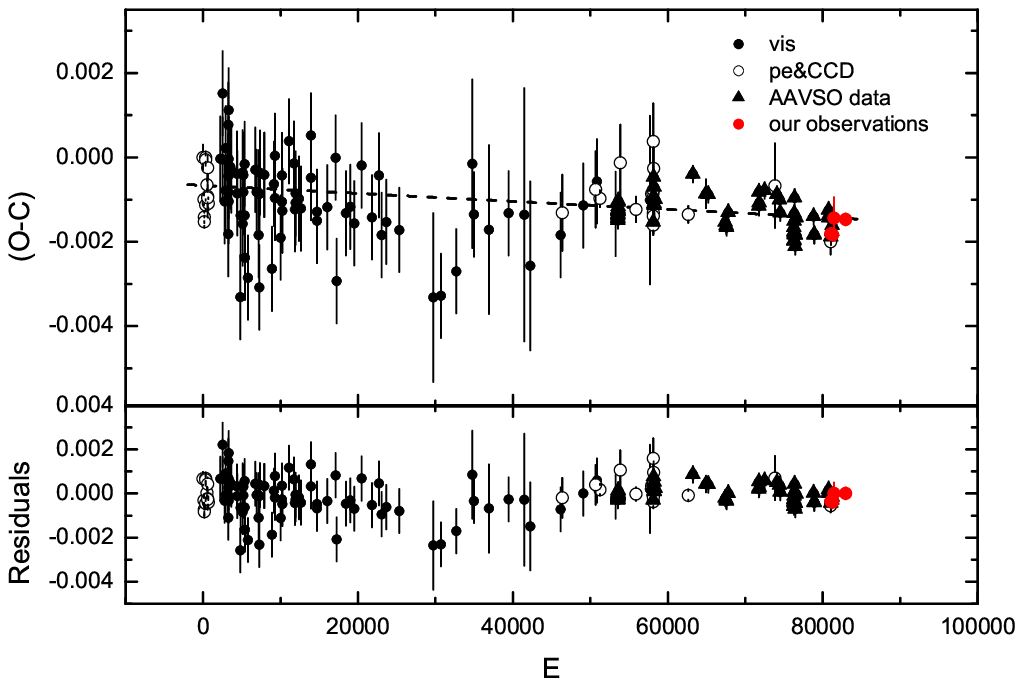}
\caption{The $O-C$ diagram of LX Ser. Upper panel shows the linear fit of the $O-C$ curve. Lower panel displays the residuals. "Vis" represents the visual data, "pe" refers to the photoelectric data, and "CCD" displays the CCD data. }
\end{center}
\end{figure}
As seen in the lower panel of Figure 3, a possible small amplitude oscillation can be extracted. Therefore, we intended to use a simultaneous linear-plus-sinusoidal function to fit the $O-C$ curve. In order to determine the period of the sinusoidal variation, the Period04 package (Lenz \& Breger 2005) was employed. The resulting Fourier spectrum is shown in Figure 4. A peak at $f=1.90386493(\pm0.06022577)\times10^{-5}$ was obtained, meaning that the period of the sinusoidal variation is 22.80$(\pm0.72)$ yr.

\begin{figure}
\begin{center}
\includegraphics[angle=0,scale=0.7]{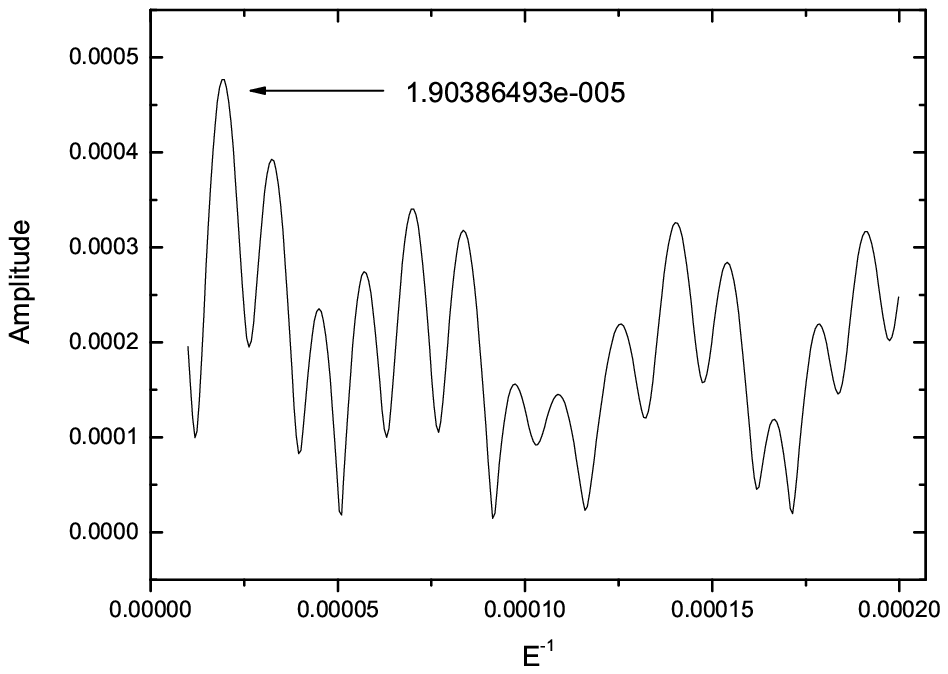}
\caption{The fourier spectrum for the orbital period variation. }
\end{center}
\end{figure}
Using the weighted least-squares method, we determined the following equation,
\begin{eqnarray}
Min.I = BJD2444293.023557(\pm0.000191) + 0^d.158432495(\pm0.000000002)\times E + \nonumber\\+ 0^d.00035(\pm0.00008) \times \sin {[0^{\circ}.00685\times E+32^{\circ}.8(\pm0.3)]}.
\end{eqnarray}
\begin{figure}
\begin{center}
\includegraphics[angle=0,scale=0.7]{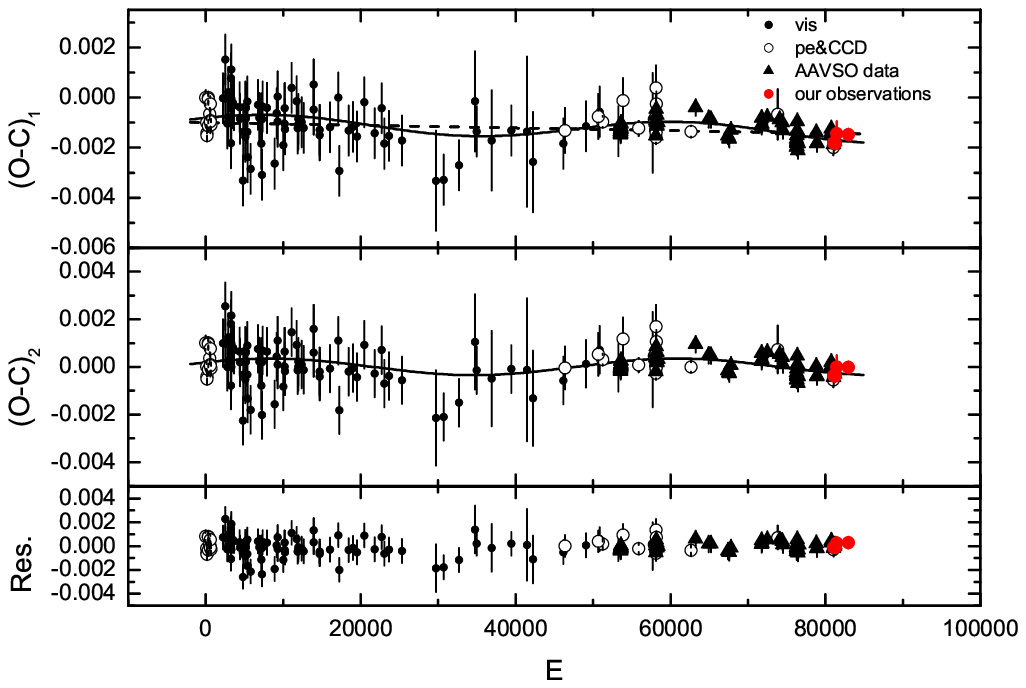}
\caption{The best fitting curve for the linear-plus-sinusoidal terms of LX Ser. }
\end{center}
\end{figure}
The variance is $\sigma_2=4.67\times10^{-7}$, which is smaller than the value of the linear fit. The $F$-test method as discussed by Pringle (1975) was used to estimate to what extent the statistical significance is
improved between the linear fit and the linear-plus-sinusoidal fit. The statistic parameter $\lambda$ is corrected to be
\begin{eqnarray}
\lambda={(\sigma_1^2-\sigma_2^2)/2\over \sigma_2^2/(n-4)}.
\end{eqnarray}
In this equation, $n$ is the number of eclipse times. We obtained $F(2,184)=19.5$, indicating that the statistical significance of the linear-plus-sinusoidal fit with respect to the linear fit is more than the 99.99\% level. The best fitting curve for the linear-plus-sinusoidal terms is displayed in the upper panel of  Figure 5. The linear term of Equation (3) represents the revision on the initial linear ephemeris (Equation 1). When the linear term was removed, the $(O-C)_2$ values are displayed in the middle panel of Figure 5. A 22.8 yr cyclic modulation with an amplitude of 0.00035 days can be seen. The residuals from Equation (3) are shown in the lower panel of Figure 5, where no regular changes can be subtracted.

\begin{table}
\scriptsize
\begin{center}
\caption{All avaiable eclipse times of LX Ser}
\begin{tabular}{llcccrrrr}\hline\hline
HJD & BJD &  Errors & Method & Cycle & O-C & (O-C)$_2^a$& (O-C)$_3^b$& References \\
2400000+&2400000+&&&&&&\\\hline
44293.02430 &	44293.02487 &	0.00030 &	pe	&0	    & 0        &  0.00067  & 0.00082  & Horne 1980                \\
44312.98580 &	44312.98637 &		--    &  pe	&126	  & -0.00100 &	-0.00031 & -0.00017 & Africano \& Klimke 1981   \\
44316.94620 &	44316.94677 &		--    &  pe	&151	  & -0.00141 &	-0.00072 & -0.00058 & Africano \& Klimke 1981   \\
44320.90690 &	44320.90747 &		--    &  pe	&176	  & -0.00152 &	-0.00083 & -0.00069 & Africano \& Klimke 1981   \\
44343.72270 &	44343.72327 &		--    &  pe	&320	  & 0.00000  &  0.00068  & 0.00081  & Africano \& Klimke 1981   \\
44344.83060 &	44344.83117 &		--    &  pe	&327	  & -0.00113 &	-0.00044 & -0.00031 & Africano \& Klimke 1981   \\
44346.89130 &	44346.89187 &		--    &  pe	&340	  & -0.00005 &	0.00063  & 0.00076  & Africano \& Klimke 1981   \\
44372.71520 &	44372.71576 &		--    &  pe	&503	  & -0.00065 &	0.00003  & 0.00016  & Africano \& Klimke 1981   \\
44384.75580 &	44384.75636 &	0.00050 &	pe	&579	  & -0.00092 &	-0.00023 & -0.00011 & Young et al. 1981         \\
44384.91420 &	44384.91476 &	0.00030 &	pe	&580	  & -0.00095 &	-0.00026 & -0.00014 & Young et al. 1981         \\
44385.86550 &	44385.86606 &	0.00030 &	pe	&586	  & -0.00025 &	0.00043  & 0.00056  & Young et al. 1981         \\
44396.79650 &	44396.79706 &		--    &  pe	&655	  & -0.00109 &	-0.00040 & -0.00028 & Africano \& Klimke 1981   \\
44644.58600 &	44644.58656 &		--    &  vis&	2219	& -0.00003 &	0.00067  & 0.00073  & BBSAG 53                  \\
44691.48200 &	44691.48255 &		--    &  vis&	2515	& -0.00005 &	0.00065  & 0.00070  & BBSAG 53                  \\
44691.64200 &	44691.64255 &		--    &  vis&	2516	& 0.00152  &  0.00222  & 0.00228  & BBSAG 53                  \\
44731.40600 &	44731.40655 &		--    &  vis&	2767	& -0.00104 &	-0.00033 & -0.00028 & BBSAG 54                  \\
44755.48800 &	44755.48855 &		--    &  vis&	2919	& -0.00078 &	-0.00007 & -0.00002 & BBSAG 54                  \\
44755.48900 &	44755.48955 &		--    &  vis&	2919	& 0.00022  &  0.00092  & 0.00097  & BBSAG 54                  \\
44808.40600 &	44808.40656 &		--    &  vis&	3253	& 0.00077  &  0.00148  & 0.00151  & BBSAG 56                  \\
44809.35400 &	44809.35456 &		--    &  vis&	3259	& -0.00182 &	-0.00110 & -0.00107 & BBSAG 56                  \\
44809.35500 &	44809.35556 &		--    &  vis&	3259	& -0.00082 &	-0.00010 & -0.00007 & BBSAG 56                  \\
44811.41500 &	44811.41556 &		--    &  vis&	3272	& -0.00044 &	0.00026  & 0.00030  & BBSAG 56                  \\
44812.36500 &	44812.36556 &		--    &  vis&	3278	& -0.00104 &	-0.00032 & -0.00029 & BBSAG 56                  \\
44812.36600 &	44812.36656 &		--    &  vis&	3278	& -0.00004 &	0.00067  & 0.00070  & BBSAG 56                  \\
44817.43700 &	44817.43756 &		--    &  vis&	3310	& 0.00112  &  0.00183  & 0.00186  & BBSAG 56                  \\
44848.33000 &	44848.33056 &		--    &  vis&	3505	& -0.00022 &	0.00049  & 0.00052  & BBSAG 56                  \\
44869.24300 &	44869.24356 &		--    &  vis&	3637	& -0.00031 &	0.00040  & 0.00043  & BBSAG 56                  \\
44987.59200 &	44987.59256 &		--    &  vis&	4384	& -0.00039 &	0.00033  & 0.00033  & BBSAG 58                  \\
44994.72100 &	44994.72156 &		--    &  vis&	4429	& -0.00085 &	-0.00012 & -0.00012 & BBSAG 58                  \\
45053.49700 &	45053.49756 &		--    &  vis&	4800	& -0.00331 &	-0.00258 & -0.00258 & BBSAG 59                  \\
45061.57900 &	45061.57956 &		--    &  vis&	4851	& -0.00137 &	-0.00064 & -0.00064 & BBSAG 60                  \\
45101.50500 &	45101.50556 &		--    &  vis&	5103	& -0.00036 &	0.00037  & 0.00036  & BBSAG 61                  \\
45104.51400 &	45104.51456 &		--    &  vis&	5122	& -0.00157 &	-0.00084 & -0.00085 & BBSAG 60                  \\
45114.49600 &	45114.49656 &		--    &  vis&	5185	& -0.00082 &	-0.00009 & -0.00010 & BBSAG 61                  \\
45115.44700 &	45115.44756 &		--    &  vis&	5191	& -0.00042 &	0.00031  & 0.00030  & BBSAG 60                  \\
45115.44700 &	45115.44756 &		--    &  vis&	5191	& -0.00042 &	0.00031  & 0.00030  & BBSAG 60                  \\
45139.52900 &	45139.52956 &		--    &  vis&	5343	& -0.00016 &	0.00057  & 0.00055  & BBSAG 61                  \\
45142.53800 &	45142.53856 &		--    &  vis&	5362	& -0.00137 &	-0.00063 & -0.00065 & BBSAG 61                  \\
45148.39900 &	45148.39956 &		--    &  vis&	5399	& -0.00238 &	-0.00164 & -0.00166 & BBSAG 61                  \\
45208.28600 &	45208.28657 &		--    &  vis&	5777	& -0.00285 &	-0.00211 & -0.00213 & BBSAG 62                  \\
45358.64100 &	45358.64157 &		--    &  vis&	6726	& -0.00029 &	0.00045  & 0.00041  & BBSAG 64                  \\
45385.57400 &	45385.57457 &		--    &  vis&	6896	& -0.00082 &	-0.00007 & -0.00011 & BBSAG 65                  \\
45432.46900 &	45432.46957 &		--    &  vis&	7192	& -0.00184 &	-0.00109 & -0.00113 & BBSAG 66                  \\
45432.47000 &	45432.47057 &		--    &  vis&	7192	& -0.00084 &	-0.00008 & -0.00013 & BBSAG 66                  \\
45439.44100 &	45439.44157 &		--    &  vis&	7236	& -0.00087 &	-0.00011 & -0.00016 & BBSAG 66                  \\
45442.44900 &	45442.44957 &		--    &  vis&	7255	& -0.00308 &	-0.00233 & -0.00237 & BBSAG 66                  \\
45459.40400 &	45459.40457 &		--    &  vis&	7362	& -0.00036 &	0.00038  & 0.00034  & BBSAG 66                  \\
45548.44300 &	45548.44358 &		--    &  vis&	7924	& -0.00041 &	0.00034  & 0.00029  & BBSAG 68                  \\
45548.44300 &	45548.44358 &		--    &  vis&	7924	& -0.00041 &	0.00034  & 0.00029  & BBSAG 68                  \\
45701.64500 &	45701.64559 &		--    &  vis&	8891	& -0.00264 &	-0.00187 & -0.00192 & BBSAG 70                  \\
45741.57200 &	45741.57259 &		--    &  vis&	9143	& -0.00063 &	0.00013  & 0.00008  & BBSAG 71                  \\
45766.60400 &	45766.60459 &		--    &  vis&	9301	& -0.00096 &	-0.00019 & -0.00024 & BBSAG 71                  \\
45766.60500 &	45766.60559 &		--    &  vis&	9301	& 0.00004  &  0.00080  & 0.00075  & BBSAG 71                  \\
45879.40700 &	45879.40759 &		--    &  vis&	10013 & -0.00190 &  -0.00112 & -0.00117 & BBSAG 72                  \\
45906.34200 &	45906.34259 &		--    &  vis&	10183 & -0.00042 &	0.00035  & 0.00030  & BBSAG 73                  \\
45908.40100 &	45908.40159 &		--    &  vis&	10196 & -0.00105 &	-0.00027 & -0.00032 & BBSAG 73                  \\
45911.41100 &	45911.41159 &		--    &  vis&	10215 & -0.00126 &	-0.00048 & -0.00053 & BBSAG 73                  \\
46046.71400 &	46046.71460 &		--    &  vis&	11069 & 0.00039  &	0.00117  & 0.00112  & BBSAG 75                  \\
46154.60600 &	46154.60660 &		--    &  vis&	11750 & -0.00014 &	0.00064  & 0.00061  & BBSAG 76                  \\
\end{tabular}
\end{center}
\end{table}

\addtocounter{table}{-1}
\begin{table}
\scriptsize
\begin{center}
\caption{ --- continued }
\begin{tabular}{llcccrrrr}\hline\hline
HJD & BJD &  Errors & Method & Cycle & O-C & (O-C)$_2^a$& (O-C)$_3^b$& References \\
2400000+&2400000+&&&&&&\\\hline
46175.51800 &	46175.51860 &		--    &  vis&	11882 & -0.00123 &	-0.00043 & -0.00047 & BBSAG 76                  \\
46183.44000 &	46183.44060 &		--    &  vis&	11932 & -0.00086 &	-0.00006 & -0.00009 & BBSAG 77                  \\
46186.45000 &	46186.45060 &		--    &  vis&	11951 & -0.00107 &	-0.00027 & -0.00031 & BBSAG 77                  \\
46252.35800 &	46252.35862 &		--    &  vis&	12367 & -0.00098 &	-0.00018 & -0.00021 & BBSAG 77                  \\
46296.40200 &	46296.40262 &		--    &  vis&	12645 & -0.00121 &	-0.00041 & -0.00043 & BBSAG 78                  \\
46497.61200 &	46497.61262 &		--    &  vis&	13915 & -0.00048 &	0.00032  & 0.00033  & BBSAG 79                  \\
46497.61300 &	46497.61362 &		--    &  vis&	13915 & 0.00052  &  0.00132  & 0.00133  & BBSAG 79                  \\
46622.45600 &	46622.45663 &		--    &  vis&	14703 & -0.00129 &	-0.00047 & -0.00045 & BBSAG 80                  \\
46625.46600 &	46625.46663 &		--    &  vis&	14722 & -0.00150 &  -0.00068 & -0.00066 & BBSAG 81                  \\
46831.58700 &	46831.58764 &		--    &  vis&	16023 & -0.00118 &	-0.00034 & -0.00029 & BBSAG 82                  \\
47003.32900 &	47003.32964 &		--    &  vis&	17107 & -0.00001 &	0.00083  & 0.00092  & BBSAG 84                  \\
47023.44700 &	47023.44764 &		--    &  vis&	17234 & -0.00293 &	-0.00208 & -0.00199 & BBSAG 85                  \\
47212.61700 &	47212.61765 &		--    &  vis&	18428 & -0.00132 &	-0.00046 & -0.00033 & BBSAG 87                  \\
47304.50800 &	47304.50866 &		--    &  vis&	19008 & -0.00117 &	-0.00031 & -0.00015 & BBSAG 88                  \\
47383.40700 &	47383.40766 &		--    &  vis&	19506 & -0.00156 &	-0.00069 & -0.00052 & BBSAG 89                  \\
47535.66200 &	47535.66266 &		--    &  vis&	20467 & -0.00019 &	0.00068  & 0.00089  & BBSAG 90                  \\
47746.37600 &	47746.37665 &		--    &  vis&	21797 & -0.00142 &	-0.00053 & -0.00027 & BBSAG 92                  \\
47890.70900 &	47890.70965 &		--    &  vis&	22708 & -0.00042 &	0.00046  & 0.00076  & BBSAG 93                  \\
47942.51500 &	47942.51566 &		--    &  vis&	23035 & -0.00184 &	-0.00094 & -0.00063 & BBSAG 94                  \\
48039.47600 &	48039.47666 &		--    &  vis&	23647 & -0.00153 &	-0.00062 & -0.00029 & BBSAG 95                  \\
48306.59300 &	48306.59367 &	0.00100 &	vis	&25333  &-0.00172 	& -0.00080 & -0.00041 & BBSAG 97                   \\
49004.64500 &	49004.64566 &	0.00200 &	vis	&29739  &-0.00332 	& -0.00236 & -0.00185 & BBSAG 103                  \\
49158.48300 &	49158.48366 &	0.00100 &	vis	&30710  &-0.00328 	& -0.00231 & -0.00178 & BBSAG 104                  \\
49475.50700 &	49475.50767 &		--    &  vis&	32711 & -0.00270 &  -0.00171 & -0.00116 & BBSAG 106                 \\
49799.50400 &	49799.50469 &	0.00200 &	vis	&34756  &-0.00015 	& 0.00085  & 0.00140  & BBSAG 108                  \\
49836.57600 &	49836.57669 &	0.00100 &	vis	&34990  &-0.00135 	& -0.00034 & 0.00020  & BBSAG 109                  \\
50139.65700 &	50139.65771 &	0.00200 &	vis	&36903  &-0.00171 	& -0.00068 & -0.00015 & BBSAG 111                  \\
50539.54100 &	50539.54172 &	0.00100 &	vis	&39427  &-0.00132 	& -0.00026 & 0.00020  & BBSAG 114                  \\
50864.48600 &	50864.48675 &	0.00300 &	vis	&41478  &-0.00136 	& -0.00028 & 0.00011  & BBSAG 117                  \\
50988.37900 &	50988.37975 &	0.00200 &	vis	&42260  &-0.00257 	& -0.00149 & -0.00111 & BBSAG 118                  \\
51603.57310 &	51603.57388 &		--    &  vis&	46143 & -0.00184 &	-0.00072 & -0.00051 & BRNO 32                   \\
51641.43900 &	51641.43978 &	0.00090 &	ccd	&46382  &-0.00131 	& -0.00019 & 0.00001  & OEJV 74                    \\
52072.53400 &	52072.53477 &		--    &  vis&	49103 & -0.00114 &	0        & 0.00007  & BBSAG 125                 \\
52320.63969 &	52320.64046 &	0.00090 &	ccd	&50669  &-0.00075 	& 0.00040  & 0.00040  & OEJV 74                    \\
52348.52400 &	52348.52476 &		--    &  vis&	50845 & -0.00057 &	0.00059  & 0.00059  & BBSAG 127                 \\
52410.47070 &	52410.47146 &	0.00020 &	pe	&51236  &-0.00097 	& 0.00018  & 0.00017  & Agerer \& Hubscher 2003    \\
52730.50400 &	52730.50475 &	0.00100 &	vis	&53256  &-0.00134 	& -0.00015 & -0.00025 & Diethelm 2003              \\
52777.87530 &	52777.87605 &	0.00008 &	pe	&53555  &-0.00135 	& -0.00016 & -0.00027 & This paper                 \\
52778.82581 &	52778.82656 &	0.00010 &	pe	&53561  &-0.00144 	& -0.00025 & -0.00036 & This paper                 \\
52779.77639 &	52779.77714 &	0.00008 &	pe	&53567  &-0.00145 	& -0.00026 & -0.00037 & This paper                 \\
52779.93490 &	52779.93565 &	0.00012 &	pe	&53568  &-0.00138 	& -0.00019 & -0.00030 & This paper                 \\
52780.72718 &	52780.72793 &	0.00015 &	pe	&53573  &-0.00126 	& -0.00007 & -0.00018 & This paper                 \\
52780.88538 &	52780.88613 &	0.00013 &	pe	&53574  &-0.00149 	& -0.00030 & -0.00041 & This paper                 \\
52781.83613 &	52781.83688 &	0.00012 &	pe	&53580  &-0.00134 	& -0.00015 & -0.00026 & This paper                 \\
52782.78677 &	52782.78752 &	0.00012 &	pe	&53586  &-0.00129 	& -0.00010 & -0.00021 & This paper                 \\
52782.94532 &	52782.94607 &	0.00013 &	pe	&53587  &-0.00117 	& 0.00001  & -0.00009 & This paper                 \\
52783.73740 &	52783.73815 &	0.00009 &	pe	&53592  &-0.00126 	& -0.00007 & -0.00018 & This paper                 \\
52786.74781 &	52786.74856 &	0.00013 &	pe	&53611  &-0.00106 	& 0.00012  & 0.00001  & This paper                 \\
52786.90609 &	52786.90684 &	0.00009 &	pe	&53612  &-0.00122 	& -0.00003 & -0.00014 & This paper                 \\
52787.85687 &	52787.85762 &	0.00010 &	pe	&53618  &-0.00103 	& 0.00015  & 0.00004  & This paper                 \\
52828.41650 &	52828.41725 &	0.00090 &	ccd	&53874  &-0.00012 	& 0.00106  & 0.00094  & Zejda 2004                 \\
53146.23100 &	53146.23174 &	0.00030 &	ccd	&55880  &-0.00123 	& -0.00002 & -0.00020 & Krajci 2005                \\
53436.63800 &	53436.63873 &	0.00200 &	vis	&57713  &-0.00101 	& 0.00021  & -0.00001 & OEJV 3                     \\
53465.63130 &	53465.63203 &	0.00030 &	ccd	&57896  &-0.00086 	& 0.00036  & 0.00013  & Zejda et al. 2006          \\
53498.26840 &	53498.26913 &		--    &  ccd&	58102 & -0.00085 &	0.00037  & 0.00013  & VSOLJ 44                  \\
53500.48638 &	53500.48711 &	0.00011 &	pe	&58116  &-0.00093 	& 0.00029  & 0.00005  & This paper                 \\
53500.80300 &	53500.80373 &	0.00020 &	ccd	&58118  &-0.00118 	& 0.00004  & -0.00019 & Krajci 2006                \\
53500.96130 &	53500.96203 &	0.00030 &	ccd	&58119  &-0.00131 	& -0.00008 & -0.00032 & Krajci 2006                \\
\end{tabular}
\end{center}
\end{table}

\addtocounter{table}{-1}
\begin{table}
\scriptsize
\begin{center}
\caption{ --- continued }
\begin{tabular}{llcccrrrr}\hline\hline
HJD & BJD &  Errors & Method & Cycle & O-C & (O-C)$_2^a$& (O-C)$_3^b$& References \\
2400000+&2400000+&&&&&&\\\hline
53501.75400 &	53501.75473 &	0.00060 &	ccd	&58124  &-0.00077 	& 0.00045  & 0.00021  & Krajci 2006                \\
53501.91160 &	53501.91233 &	0.00020 &	ccd	&58125  &-0.00160  &  -0.00037 & -0.00061 & Krajci 2006                \\
53502.54540 &	53502.54613 &	0.00031 &	pe	&58129  &-0.00153 	& -0.00030 & -0.00054 & This paper                 \\
53504.76360 &	53504.76433 &	0.00020 &	ccd	&58143  &-0.00139 	& -0.00016 & -0.00040 & Krajci 2006                \\
53504.92380 &	53504.92453 &	0.00090 &	ccd	&58144  &0.00038   &  0.00160  & 0.00136  & Krajci 2006                \\
53506.50728 &	53506.50801 &	0.00033 &	pe	&58154  &-0.00046 	& 0.00076  & 0.00052  & This paper                 \\
53510.46830 &	53510.46903 &	0.00050 &	ccd	&58179  &-0.00026 	& 0.00096  & 0.00073  & Hubscher et al. 2005       \\
53514.42838 &	53514.42911 &	0.00037 &	pe	&58204  &-0.00099 	& 0.00023  & 0.00000  & This paper                 \\
53516.48800 &	53516.48873 &	0.00035 &	pe	&58217  &-0.00099 	& 0.00023  & 0.00000  & This paper                 \\
53519.49811 &	53519.49884 &	0.00042 &	pe	&58236  &-0.00110  &  0.00012  & -0.00011 & This paper                 \\
53521.39970 &	53521.40043 &	0.00039 &	pe	&58248  &-0.00070  &  0.00052  & 0.00028  & This paper                 \\
53521.55769 &	53521.55842 &	0.00040 &	pe	&58249  &-0.00114 	& 0.00008  & -0.00015 & This paper                 \\
53541.52033 &	53541.52106 &	0.00012 &	pe	&58375  &-0.00100  &  0.00023  & -0.00001 & This paper                 \\
54218.34361 &	54218.34434 &	0.00020 &	ccd	&62647  &-0.00135 	& -0.00008 & -0.00034 & OEJV 107                   \\
54316.41428 &	54316.41502 &	0.00008 &	ccd	&63266  &-0.00040  &  0.00087  & 0.00061  & This paper                 \\
54580.52079 &	54580.52153 &	0.00034 &	ccd	&64933  &-0.00086 	& 0.00043  & 0.00018  & This paper                 \\
54628.52582 &	54628.52656 &	0.00013 &	ccd	&65236  &-0.00088 	& 0.00041  & 0.00017  & This paper                 \\
54976.44292 &	54976.44368 &	0.00014 &	ccd	&67432  &-0.00153 	& -0.00021 & -0.00040 & This paper                 \\
54994.50413 &	54994.50489 &	0.00006 &	ccd	&67546  &-0.00162 	& -0.00030 & -0.00049 & This paper                 \\
55001.47512 &	55001.47588 &	0.00007 &	ccd	&67590  &-0.00166 	& -0.00034 & -0.00053 & This paper                 \\
55037.43966 &	55037.44042 &	0.00017 &	ccd	&67817  &-0.00130  &  0.00001  & -0.00016 & This paper                 \\
55662.45633 &	55662.45711 &	0.00016 &	ccd	&71762  &-0.00082 	& 0.00053  & 0.00047  & This paper                 \\
55663.40660 &	55663.40738 &	0.00022 &	ccd	&71768  &-0.00115 	& 0.00020  & 0.00015  & This paper                 \\
55672.43733 &	55672.43811 &	0.00016 &	ccd	&71825  &-0.00107 	& 0.00028  & 0.00023  & This paper                 \\
55778.42896 &	55778.42974 &	0.00011 &	ccd	&72494  &-0.00078 	& 0.00058  & 0.00055  & This paper                 \\
55988.66900 &	55988.66978 &	0.00100 &	ccd	&73821  &-0.00067 	& 0.00070  & 0.00072  & OEJV 147                   \\
56028.43535 &	56028.43613 &	0.00010 &	ccd	&74072  &-0.00088 	& 0.00050  & 0.00053  & This paper                 \\
56088.48114 &	56088.48192 &	0.00009 &	ccd	&74451  &-0.00100  &  0.00038  & 0.00042  & This paper                 \\
56101.63073 &	56101.63151 &	0.00012 &	ccd	&74534  &-0.00131 	& 0.00007  & 0.00012  & This paper                 \\
56378.88725 &	56378.88804 &	0.00014 &	ccd	&76284  &-0.00166 	& -0.00026 & -0.00014 & This paper                 \\
56381.89715 &	56381.89794 &	0.00019 &	ccd	&76303  &-0.00198 	& -0.00058 & -0.00046 & This paper                 \\
56383.79856 &	56383.79935 &	0.00029 &	ccd	&76315  &-0.00176 	& -0.00036 & -0.00024 & This paper                 \\
56384.43209 &	56384.43288 &	0.00015 &	ccd	&76319  &-0.00196 	& -0.00056 & -0.00044 & This paper                 \\
56384.59069 &	56384.59148 &	0.00015 &	ccd	&76320  &-0.00179 	& -0.00039 & -0.00027 & This paper                 \\
56384.74908 &	56384.74987 &	0.00016 &	ccd	&76321  &-0.00183 	& -0.00043 & -0.00031 & This paper                 \\
56385.85843 &	56385.85922 &	0.00020 &	ccd	&76328  &-0.00151 	& -0.00011 & 0.00000  & This paper                 \\
56386.80920 &	56386.80999 &	0.00017 &	ccd	&76334  &-0.00133 	& 0.00006  & 0.00018  & This paper                 \\
56389.50214 &	56389.50293 &	0.00011 &	ccd	&76351  &-0.00175 	& -0.00035 & -0.00023 & This paper                 \\
56389.66136 &	56389.66215 &	0.00021 &	ccd	&76352  &-0.00096 	& 0.00043  & 0.00055  & This paper                 \\
56403.44384 &	56403.44463 &	0.00008 &	ccd	&76439  &-0.00211 	& -0.00071 & -0.00058 & This paper                 \\
56410.41514 &	56410.41593 &	0.00016 &	ccd	&76483  &-0.00184 	& -0.00044 & -0.00031 & This paper                 \\
56427.68469 &	56427.68548 &	0.00020 &	ccd	&76592  &-0.00143 	& -0.00002 & 0.00009  & This paper                 \\
56782.41510 &	56782.41588 &	0.00006 &	ccd	&78831  &-0.00140  &  0.00002  & 0.00022  & This paper                 \\
56792.39590 &	56792.39668 &	0.00020 &	ccd	&78894  &-0.00184 	& -0.00041 & -0.00021 & This paper                 \\
56798.41636 &	56798.41714 &	0.00006 &	ccd	&78932  &-0.00182 	& -0.00039 & -0.00019 & This paper                 \\
57091.67550 &	57091.67627 &	0.00007 &	ccd	&80783  &-0.00125 	& 0.00018  & 0.00044  & This paper                 \\
57094.68559 &	57094.68636 &	0.00021 &	ccd	&80802  &-0.00137 	& 0.00007  & 0.00033  & This paper                 \\
57097.69575 &	57097.69652 &	0.00005 &	ccd	&80821  &-0.00143 	& 0.00001  & 0.00027  & This paper                 \\
57132.39190 &	57132.39267 &	0.00030 &	pe	&81040  &-0.00200  &  -0.00055 & -0.00029 & Hubscher 2016              \\
57134.45164 &	57134.45241 &	0.00004 &	ccd	&81053  &-0.00188 	& -0.00043 & -0.00017 & This paper                 \\
57135.40230 &	57135.40307 &	0.00020 &	pe	&81059  &-0.00182 	& -0.00037 & -0.00011 & Hubscher 2016              \\
57137.30350 &	57137.30427 &	0.00007 &	ccd	&81071  &-0.00181 	& -0.00036 & -0.00010 & This paper                 \\
57159.48411 &	57159.48488 &	0.00005 &	ccd	&81211  &-0.00175 	& -0.00030 & -0.00003 & This paper                 \\
57163.44506 &	57163.44583 &	0.00013 &	ccd	&81236  &-0.00161 	& -0.00016 & 0.00010  & This paper                 \\
57173.26766 &	57173.26843 &	0.00015 &	ccd	&81298  &-0.00183 	& -0.00038 & -0.00011 & This paper                 \\
57200.04314 &	57200.04391 &	0.00049 &	ccd	&81467  &-0.00144 	& 0        & 0.00028  & This paper                 \\
57439.27618 &	57439.27695 &	0.00004 &	ccd	&82977  &-0.00147 	& 0        & 0.00029  & This paper                 \\\hline
\end{tabular}
\end{center}
$^a$ (O-C)$_2$ means the residuals for Equation (2), $^b$ (O-C)$_3$ means the residuals for Equation (3)
\end{table}

\section {Discussion}	
\subsection{The cyclic orbital period variation}
Based on the investigation of the $O-C$ diagram, we found that the orbital period of LX Ser shows a cyclic modulation with a period of $P_3=22.8$ yr and an amplitude of $A_3=0.00035$ days. The cyclic change of the $O-C$ diagram can be explained either by the Applegate mechanism (Applegate 1992) due to the magnetic activity of the red dwarf component or by the light travel time effect due to a tertiary companion.

The Applegate mechanism was proposed according to the conclusion determined by Hall (1989) who analyzed the orbital period variations of 101 Algols. Hall (1989) found that all the binary systems which show cyclic period changes have late spectral type secondary components. However, a recently statistical study by Liao \& Qian (2010) reveals that the percentages for both late type and early type binary systems that exhibit cyclic orbital period changes are similar, and they concluded that the cyclic oscillation is most likely caused by the light travel time effect via the presence of a third body.

According to Applegate mechanism, solar-like magnetic cycles would result in shape changes of the low-mass components, thus redistributing the angular momentum within the interior of the star. Then, the oblateness is changed, causing a change of the stellar quadrupole moment which consequently leads to the variation of orbital period.
Using the equation
\begin{equation}
{\Delta{P}\over P}=2\pi{O-C\over P_{mod}},
\end{equation}
where $(O-C)$ and $P_{mod}$ are the amplitude and modulation period of the cyclic oscillation of the orbital period. The fractional period change $\Delta{P}\over P$ was determined to be $2.67\times 10^{-7}$. By considering a typical mass of $M_1=0.6$ $M_\odot$ for the primary white dwarf of LX Ser, the mass of the secondary red dwarf can be calculated to be $M_2=q\times M_1=$0.3 $M_\odot$ (the mass ratio $q$ is 0.5 based on the result of Marin et al. 2007). The radius of the secondary red dwarf was estimated to be 0.38 $R_\odot$ based on Cox (2000), the binary separation $a=1.19R_\odot$ can be derived using $M_1+M_2=0.0134a^3/P^2$.
Then, according to the equation
\begin{equation}
{\Delta{P}\over P}=-9({R_2\over a})^2{\Delta Q\over M_2R_2^2},
\end{equation}
we determined the variation of the quadrupole moment of the red dwarf to be $\Delta Q=-1.21\times10^{47}$ g cm$^2$.
The secondary component of LX Ser may be a full convective star, we can use the same method that Brinkworth et al. (2006) used to calculate the energy required to produce the cyclic modulation of LX Ser. We split the whole star into an inner core (denotes as 1) and an outer shell (denotes as 2). By employing the Lane-Emden equation for an $n=1.5$ polytrope, different shell mass versus the required energy $\Delta E$ can be calculated and is displayed in Figure 6. The minimum value of the required energy is achieved to be $\Delta E_{min}=8.95\times 10^{40}$ erg when $M_s=0.224 M_\odot$. Assuming that the temperature of $T_2=3500$ K, the total radiated energy of the secondary red dwarf over the whole modulation period is also shown in Figure 6. Obviously, the Applegate mechanism has difficulty to explain the cyclic oscillation in the $O-C$ diagram.

\begin{figure}
\begin{center}
\includegraphics[angle=0,scale=0.7]{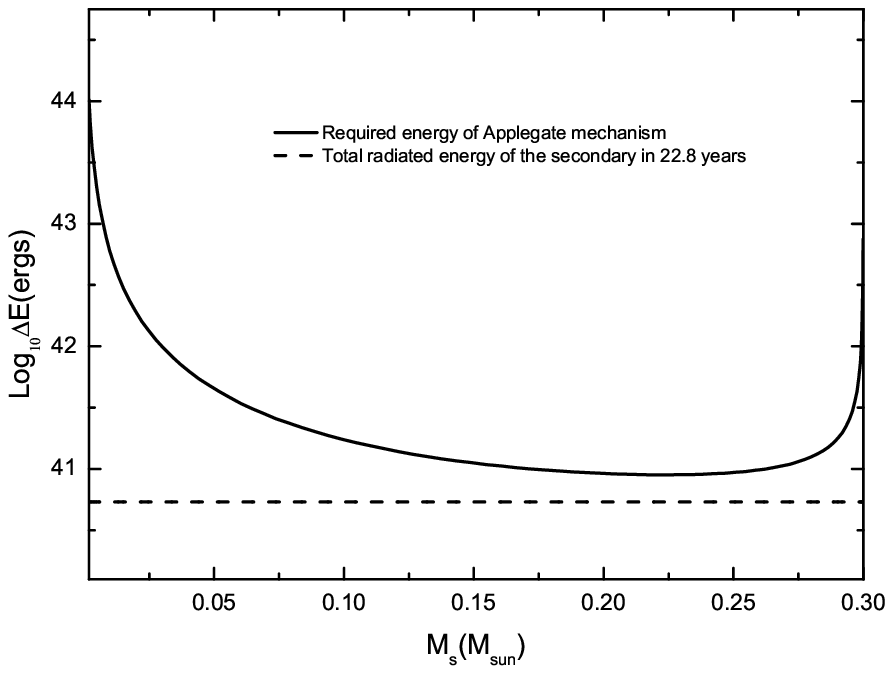}
\caption{The required energy $\Delta E$ versus different shell mass. The lowest value of required
energy is $8.95 \times 10^{40}$ erg at a shell-mass of 0.224 $M_\odot$.}
\end{center}
\end{figure}

Therefore, liking many other CVs, DQ Her (Dai \& Qian 2009), UZ For (Potter et al. 2011), AM Her (Dai et al. 2013), the plausible explanation of the cyclic modulation is the light travel time effect. Using the equation
\begin{equation}
f(m)={(m_3\sin i)^3\over (m_1+m_2+m_3)^2}={4\pi\over GP^2_3}\times(a_{12}\sin i)^3,
\end{equation}
the mass function $f(m)=4.29(\pm2.94)\times 10^{-7}\,M_\odot$ and the mass of the third companion $M_3\sin i^{\prime}=0.0071(\pm0.0027)\,M_\odot=7.4(\pm2.8) \,M_{Jup}$ were computed. Simulations by Bonnell \& Bate (1994) and Holman \& Wiegert (1999) exhibit that circumbinary planets have stable orbits and are most possibly coplanar with the central eclipsing host systems. Considering the orbital inclination of LX Ser $i=79^\circ.0$, the mass of the third body is derived to be $M_3=0.0072(\pm0.0027)\,M_\odot=7.5(\pm2.8) \,M_{Jup}$ with a separation of $9.12(\pm4.03)$ AU. Based on the theoretical analysis (e.g., Chabrier \& Baraffe 2000; Burrows et al. 2001), the upper limit mass of a planet is about $0.0140\,M_\odot \sim14.6\,M_{Jup}$, the third companion is most likely to be a giant planet.

\subsection{The stability of the possible planet}
In this subsection, we discussed the stability of the possible planet. As the mass of LX Ser is much larger than the outer planet, we can ignore it when considering the evolution of the inner CV.
The logarithmically differentiating of semi-major axis of LX Ser with respect to time gives
\begin{eqnarray}
\frac{\dot a}{a}=\frac{2\dot J_{orb}}{J_{orb}}+\frac{2(-\dot M_2)}{M_2}(1-\frac{M_2}{M_1}).
\end{eqnarray}
When assuming a constant angular momentum, the period of the orbit increases, which may impact the stability of the outer planet. However, there are many factors that influence the orbital angular momentum in close binary systems, especially in a CV system.
The dominant angular momentum losses (AML) mechanism in long-period systems($P_{orb}\ge3hrs$) is "magnetic braking", whereas short period CVs($P_{orb}\le2hrs$) are assumed to be driven by AML associated with the emission of "gravitational radiation" (Knigee \& Baraffe 2011).
Apart from the two major factors, we should also consider mass transfer, mass accretion, common-envelope evolution, wind accretion, tidal evolution and many other factors (Hurley \& Tout 2002). These factors are important in the orbital evolution of CVs. We adopted the algorithm illustrated in Hurley \& Tout (2002) and did a simulation with $M_1=0.6 M_{\odot}$, $M_2=0.3 M_{\odot}$, $P=0.^d158 $, $e=0$, $z=0.02$, where $e$ is the eccentricity and $z$ is the metallicity. The evolutions of the masses of the two components, orbital period of LX Ser are shown in Figure 7.

\begin{figure}
\vspace{0cm}\hspace{0cm}
\centering
\includegraphics[scale=0.8]{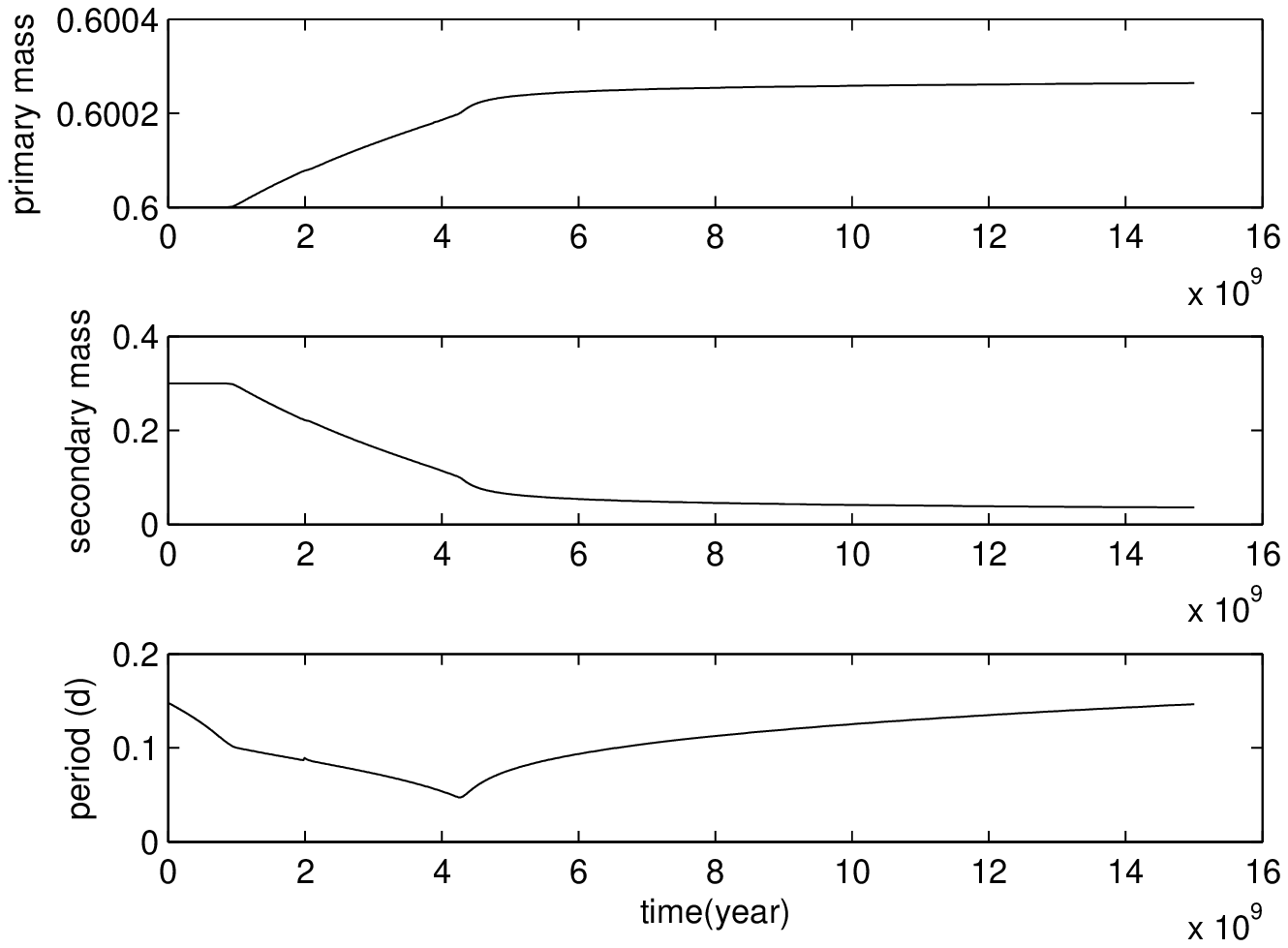}
\vspace{0cm}
\caption{Evolution of masses of the two components and period of LX Ser. }
\end{figure}

With multiple factors considered, we can see that the mass changes of the two components occur at different time and speed, which is mainly caused by wind accretion. Besides, the orbit of LX Ser contracts at the beginning of the evolution, mainly caused by gravitational mechanism. With the evolution of the secondary component, the period of the system is growing at a slow speed. Even at the universe timescale, the period is at most $0.^d16$, without much difference with its initial period. Therefore, while considering the orbital evolution of the outer planet, we can regard the inner CV as a particle with decreasing mass. The outer planet will drift slowly away from the inner CV, the system remains stable.

Now we consider whether the existence of a planet with the characteristics derived in the previous subsection is compatible with the evolution of LX Ser from a wide orbit binary to a CV. Similar consideration has been carried out in recent work of Bruch (2014). If the planet can survive the common envelope phase of the binary, the separation of the planet from the binary should be larger than the radius $R_{\rm CE}$ of the common envelope. $R_{\rm CE}$ is approximated by the radius $R_{\rm max}$ of progenitor star which evolves to be a giant star with its core mass equal to the mass of the white dwarf. According to Joss et al. (1987), the relation between the core mass and the radius is
\begin{eqnarray}
R\approx\frac{3.7\times10^3\mu^4}{1+\mu^3+1.75\mu^4}R_{\odot}.
\end{eqnarray}
With the mass of the white dwarf adopted to be 0.6 $M_{\odot}$, we can estimate the radius of the giant star in the common envelope phase to be $R_{\rm max}=1.53$ AU. As the mass loss of the primary star during the common envelope phase leads to a widening of the planetary orbit, we should compare the radius of the progenitor star to the original separation of the planet from the inner binary. According to Zhao et al. (2012), the initial-final-mass-relation for the white dwarf is
\begin{eqnarray}
M_{f}=(0.452\pm0.045)+(0.073\pm0.019)M_i.
\end{eqnarray}
Where $M_i$ is the initial mass of the white dwarf, i.e., the mass of the progenitor star, $M_f$ is the mass of the white dwarf after the common envelope phase. As $M_f$ here is $0.6\,M_{\odot}$, the mass of the progenitor is about $M_{\rm prog}=2.03\,M_{\odot}$. We assume that the orbital angular momentum of the planet is constant during the common envelope phase and the mass of the secondary star changes little. So,
\begin{eqnarray}
m_p\frac{M_{\rm prog}+M_2}{M_{\rm prog}+M_2+m_p}\sqrt{(M_{\rm prog}+M_2)a_i}=m_p\frac{M_1+M_2}{M_1+M_2+m_p}\sqrt{(M_1+M_2)a_f},
\end{eqnarray}
where $a_i$ is the original semi-major axis of the planet and $a_f$ is the current semi-major axis which is determined to be $a_f=9.12\pm4.03$ AU. At the lower limit, $a_i\simeq1.97$ AU, at the upper limit, $a_i\simeq5.08$ AU, therefore, $a_i>R_{\rm max}$ even at the lower limit. So the planet can survive the common envelope phase and have no conflicts with the evolution of the binary system.

\section {Conclusions}
The CV LX Ser was observed on four nights using the 1.0 m Cassegrain telescope at Weihai Observatory of Shandong University, and four new eclipsing times with high precision were determined. Based on the data from AAVSO International Data base, 58 eclipse times were redetermined. Further more, combining all the eclipse times from O-C Gateway, we analyzed the $O-C$ behavior of LX Ser. The $O-C$ diagram shows a cyclic variation with an amplitude of 0.00035 days and period of 22.8 yr.

For the cyclic variation in the $O-C$ diagram, both of the Applegate mechanism and the light travel time effect are considered. The energy required to produce the cyclic modulation of LX Ser was calculated, we found that the minimum value of the required energy is nearly two times of the total radiated energy of the secondary red dwarf over the whole modulation period. Therefore, the Applegate mechanism is too feeble to explain the cyclic modulation. According to the investigation of the light travel time effect, a giant planet with a mass of $7.5 \,M_{Jup}$ at a distance of 9.12 AU was explored. Exoplanets can survive almost anywhere based on the analysis during the last 24 years (e.g., Wolszczan \& Frail 1992; Silvotti et al. 2007; Qian et al. 2009, 2012). The existence of the giant planet orbiting the CV LX Ser is reasonable. By considering all the factors, we analyzed the the stability of the multiple systems and we found that the outer planet will move slowly away from the centre eclipsing CV and the system will stay stable. We also considered whether the existence of the giant planet is compatible with the evolution of LX Ser from a wide orbit binary to a CV and we determined that the planet can survive the common envelope phase and have no conflicts with the evolution of the binary system.
In the future, more eclipse times with high precision are needed to confirm the possibility of the giant planet.

\acknowledgments
This work is partly supported by the National
Natural Science Foundation of China and the Chinese Academy of Sciences joint fund on astronomy (No. U1431105) and by the National Natural Science Foundation of China (No. 11333002), and by the Natural Science Foundation of Shandong Province (No. ZR2014AQ019), and by Young Scholars Program of Shandong University, Weihai (No. 2016WHWLJH07), and by the Open Research Program of Key Laboratory for the Structure and Evolution of Celestial Objects (No. OP201303).
The authors thanks the AAVSO International Database very much for the observations of LX Ser. The data used in this paper were contributed by worldwide observers as follows: Bruno, Alain; Carreno, Alfonso; Cook, Michael; Darriba Martinez, Adolfo; Bubovsky, Pavol; Foster, James; Gomez, Tomas; Graham, Keith; Gualdoni, Carlo; James, Robert; Macdonald, Walter; Menzise, Kenneth; Poyner, Gary; Roe, James. Many thanks to the referee for the helpful comments and suggestions that helped to greatly improve this paper.

\end{document}